\documentclass{article}
\usepackage{authblk}
\usepackage{natbib}
\usepackage{apalike}
\usepackage{graphicx}
\usepackage{amssymb}

\title{Response statistics dissect the contributions of different sources of variability to \\population activity in V1}

\author{Mih\'aly B\'anyai} 
\author{Zsombor Koman}
\author{Gerg\H{o} Orb\'an}
\affil{Computational Systems Neuroscience Lab, \\MTA Wigner Research Centre for Physics, Budapest, Hungary}
\date{}

\begin{document}
\maketitle

\begin{abstract}
Response variability, as measured by fluctuating responses upon repeated performance of trials, is a major component of neural responses, and its characterization is key to interpret high dimensional population recordings. Response variability and covariability display predictable chan\-ges upon changes in stimulus and cognitive or behavioral state, providing an opportunity to test the predictive power of models of neural variability. Still, there is little agreement on which model to use as a building block for population-level analyses, and models of variability are often treated as a subject of choice. We investigate two competing models, the Doubly Stochastic Poisson (DSP) model assuming stochasticity at spike generation, and the Rectified Gaussian (RG) model that traces variability back to membrane potential variance, to analyze stimulus-dependent modulation of response statistics. Using a model of a pair of neurons, we demonstrate that the two models predict similar single-cell statistics. However, DSP and RG models have contradicting predictions on the joint statistics of spiking responses. In order to test the models against data, we build a population model to simulate stimulus change-related modulations in response statistics. We use unit recordings from the primary visual cortex of monkeys to show that while model predictions for variance are qualitatively similar to experimental data, only the RG model's predictions are compatible with joint statistics. These results suggest that models using Poisson-like variability might fail to capture important properties of response statistics. We argue that membrane potential-level modelling of stochasticity provides an efficient strategy to model correlations.  
\end{abstract}

\section*{Significance statement}
Variability and covariability in neural activity are some of the most puzzling aspects of the otherwise impressively effective neural computations. For efficient decoding and prediction, models of information encoding in neural populations hinge upon an appropriate model of neural variability. Our work shows that stimulus-dependent changes in the pairwise statistics but not in single-cell statistics can differentiate between two widely used models of neuronal variability. Importantly, contrasting model predictions with neuronal data provides hints on the sources of noise in neural spiking. These findings provide essential constraints on statistical models of population activity.

\section{Introduction}
Variability in the nervous system is ubiquitous and affects perception, decision making, and motor control \citep{Faisal:2008cp}. Characterizing the properties \citep{Ecker:2014cl, Goris:2014jg, Kohn:2005um, Lin:2015dw} and identifying the sources \citep{Renart:2014dr} of this variability is critical for understanding the computations taking place in the nervous system. In particular, in the visual system, understanding how cellular, network, or bottom-up processes contribute to variability helps to assess the interaction between the stimulus, internal states of the brain and possible noise. In order to tackle variability, the goal is to be able to consistently predict response statistics of neurons beyond the mean response under different conditions. 

The overwhelming majority of data recorded with a focus on the assessment of response variability comes from recordings of spiking activity of neurons \citep{Churchland:2010he}. Spike count measurements are characterized by a distinctive pattern that relates variability to firing rates: there seems to be a tendency that spike count variability grows linearly with mean firing rate \citep{Tolhurst:1981ju}. This form of single-cell statistics is central to many theories of coding \citep{Churchland:2011hd, Ecker:2016hc, Jazayeri:2006fk, Ma:2014in, Simoncelli:2004ue, Pillow:2007wh, Froudarakis:2014fs}: the resemblance of this statistics to the characteristics of the Poisson process led to a simple statistical model of variability, which assumes that variability arises as a consequence of a renewal process with independent spikes sampled in finite time windows with a given expected value but independent across time windows. 

While Poisson spiking statistics seems to be a sound model of variability, it suggests a very specific assumption on the source of noise in neural responses: stochasticity is assumed to arise at the level of spike generation. Variability can arise at different levels of cellular and network dynamics: at the level of membrane potential (for instance because of channel noise), at the level synaptic transmission, at the level of uncontrolled variables either in the stimulus, in the movements of the animal, or internal to the processing (e.g. attention, or top-down processing) \citep{Ruff:2014fa}. Recent studies have focused on identifying different sources of response variability by distinguishing the contributions of Poisson-like stochasticity and additional fluctuations that further increase variability  \citep{Goris:2014jg, Churchland:2011hd}. 

The assumption that it is a stochastic process in spike generation that underlies spiking variability has been challenged by the highly reliable spike generation mechanisms found in vitro \citep{Mainen:1995uz}. An alternative approach, the Rectified Gaussian model (RG), which is compatible with this result but can also account for the characteristic mean-variance relationship of spike count statistics, has been proposed by \citep{Carandini:2004ee}. According to this approach, firing rate nonlinearity, the mapping between membrane potential to firing rate, by itself can achieve the scaling of spike count variance with spike count mean. As opposed to a Poisson-like account, spiking variability in the RG model originates from variability present at the level of membrane potentials. A linear relationship between spike count mean and variance is achieved by increased stretching of the normally distributed membrane potential by the convex firing rate nonlinearity when the mean membrane potential is higher. This model has also proven successful in accounting for patterns in mean spiking responses of V1 neurons \citep{Finn:2007hc}.

Stimulus-dependence of response mean and variance, often characterized by the Fano factor, provides important insights into the forms of stochasticity responsible for response variability. Going beyond single-cell response statistics by analyzing joint statistics of the responses of multiple neurons can provide the constraints necessary to dissect possible mechanisms responsible for spiking variability. Collective changes, or correlations in their simplest form, have recently gained intense attention \citep{Ecker:2010dn, Cohen:2011eh}. Studies have demonstrated that bottom-up \citep{Kohn:2005um}, top-down \citep{Ruff:2014fa, Haefner:2016vr} and even cellular \citep{deLaRocha:2007go} factors can affect spike count correlations and others have revealed that both multiplicative and additive forms of correlations contribute to these changes \citep{Lin:2015dw}. It is unclear, however, how well alternative models of spike count variability can predict the patterns in response correlations as well as the patterns in mean responses and response variance. 

In this study we set out to contrast competing approaches proposed for describing spike count variability and use their predictions on response statistics. We define the Doubly stochastic Poisson (DSP) model and the Rectified Gaussian model to analyze the relationship between membrane potential statistics and spike count statistics. In particular, we focus on the joint spiking statistics of a pair of neurons to demonstrate a dissociation between the two models based on changes in spike count correlation resulting from changes in membrane potential statistics. Using these analyses as a starting point, we simulate stimulus change-related modulation of spike count statistics for both of the models in a population of model neurons and compare the predictions to orientation-dependent and contrast-dependent changes in the activity statistics of extracellularly recorded V1 neurons in awake monkeys. We argue that changes in stimulus attributes give rise to distinctive patterns in response correlations that are compatible with the Rectified Gaussian model but contradict the Doubly Stochastic Poisson model.

\section{Materials and Methods}

\subsection{Model of membrane potential responses}

We model the responses of simple cells of the primary visual cortex at the level of the membrane potentials. The membrane potential of a simple cells shows systematic variations with changes in the stimulus. The most widely studied of these changes, the orientation dependence was considered as changes in the trial-by-trial mean of the membrane potential response. Trial-averaged responses provide an incomplete description of the membrane potential statistics since there are both within-trial fluctuation and across trial fluctuations \citep{Tomko:1974ul}. The source of these fluctuations are not considered. A neuron is characterized by the probability distribution of membrane potential values at each time point within a trial, which random samples were drawn from. Importantly, besides stimulus-driven covariations in membrane potential responses of multiple neurons, stimulus-independent, so called noise covariances are also characteristic features of cortical neurons \citep{Yu:2010p314}. As a consequence, a population of neurons in the visual cortex is also characterized by the joint probability distribution of the membrane potentials. Similar to systematic variations in trial-averaged mean membrane potential responses, membrane potential variances are characterized by systematic changes \citep{Finn:2007hc}, which we also take into account in the model.

Instead of aiming for a complete description of such probability distributions, we focus on the second-order statistics of neural responses. This is motivated by two considerations: (i) experimental designs are typically limited in terms of the number of trials, rendering higher-order joint statistics of neurons hard to estimate (ii) pairwise statistics are the simplest measure of population activity going beyond individual cell response properties, and are already able to capture definitive signatures of population-level cortical computation \citep{Karklin:2009hl, Haefner:2013gh}. 

The second-order statistics of a population of $N$ units with temporal dynamics is completely characterized by $N$ autocorrelation functions and $N(N-1) / 2$ cross-correlation functions, specifying the linear dependence between each pair at every timescale \citep{MorenoBote:2008gg}. Based on the typical support of the autocorrelation function of membrane potentials, we chose to examine interactions on the scale of 20 ms \citep{Azouz:1999wj}. Using this, we can simplify the above picture greatly by specifying a single membrane potential value in each 20 ms bin, assuming that there is no temporal dependence on longer timescales, and omitting variability on shorter ones. Thus, we can consider membrane potentials to be sampled from a Gaussian distribution at each time bin $b$ independently:

\begin{equation}\label{eq:mp_gauss}
\mathbf{u_b} \sim \mathcal{N}(\mathbf{u_b}; \mathbf{\mu},\mathbf{C_{mp}}).
\end{equation}
The mean of the distribution, $\mu \in \mathbb{R}^N$, determines the trial-averaged membrane potential level of each neuron, and the sequences of individual samples give rise to within-trial variability. 

Since the focus of the study is to understand the implications of membrane potential fluctuations on spiking statistics, we do not seek to find a match between membrane potential recordings and the model. Rather, membrane potential statistics is assessed in terms of its consequences on spiking statistics.

\subsection{Doubly Stochastic Poisson spike generation model}

Intensity of spike responses of neurons is determined by the instantaneous firing rate function. This mapping from membrane potential to firing rate is achieved by the firing rate nonlinearity, for which we take a parametric form from the literature \citep{Carandini:2004ee}, parametrized identically for each neuron:

\begin{equation}\label{eq:firing_rate}
\mathbf{r}_b = k \left[ \mathbf{u}_b - V_{th} \right]_{+}^\beta.
\end{equation}
$V_{th}$ denotes the membrane potential threshold under which the firing rate is zero. Above the threshold the firing rate is a power law function with exponent $\beta$, where $\beta=1$ corresponds to a linear mapping. The rate is mapped to the Hertz scale using a gain parameter, $k$, which indicates how many spikes are to be expected within a single time bin (which is always 20 ms in our study). Using this mapping, we obtain an instantaneous rate in every time bin, which serves as an intermediate quantity between membrane potentials and spikes. The parameters in the rate model are chosen to be typical to the primary visual cortex based on Carandini, 2004. The values used throughout this paper are $V_{th} = 0$ (the threshold only contributes to the rate through the difference with the membrane potential, thus we need to choose $\mathbf{\mu}$ and  $V_{th}$ together to produce realistic membrane potential dynamics), $\beta = 1.4$, and $k = 0.4$, which together with the 20 ms time bin corresponds to a rate of 20 Hz, typically observed in the primary visual cortex in response to a high contrast stimulus of parameters preferred by the tuning curve of the cell \citep{Finn:2007hc}.  

The firing rate nonlinearity establishes the link between average membrane potential and average firing rate. A widely used approach \citep{Gur:1997ur} assumes that the firing rate determines the probability with which a spike is generated in any particular time window. When these probabilities are independent across time we formally obtain the Poisson process. In this model, spike counts are sampled from a Poisson distribution, parametrized by the instantaneous rate, independently for each neuron $n$.

\begin{equation}\label{eq:poisson_sampling}
s_b^n \sim \mathrm{Poisson}(s_b^n; r_b^n)
\end{equation}

In summary, the Poisson model of spiking, the Doubly Stochastic Poisson model (DSP), relies on two sources of variability: 1, membrane potentials are stochastically generated and these samples are correlated to represent covariability of neuronal responses; 2, a second source of stochasticity comes from the generation of spikes which introduces noise that is independent across neurons.

\subsection{Rectified Gaussian model}

The DSP model assumes that the firing rate defines the probability with which spikes are generated in a given time window. Inspired by the relatively little stochasticity found in sensory neurons in vitro \citep{Mainen:1995uz}, an alternative model can be formulated, which assumes a deterministic process for spike generation \citep{Carandini:2004ee}. Self-consistency requires that the number of spikes with constant stimulus on average is proportional to the firing rate. A model that assumes no further source of variability but fulfils the self consistency criterion can be formulated by integrating the firing rate over time and generating spikes whenever the integral crosses integer values. If we consider the rate to be constant within a time bin $b$, scaled appropriately with the base rate parameter $k$, the integral becomes a finite sum, for neuron $n$:

\begin{equation}\label{eq:sc_integration}
s_t^n = \lfloor \sum_{b=1}^{\tau} r_b^n \rfloor
\end{equation}
The spiking model defined this way, the Rectified Gaussian model (RG), is formally equivalent to an integrate-and-fire neuron model without refractory period with the addition of the firing rate nonlinearity. The principal source of variability in the RG model is coming from the variability present in the membrane potentials. An additional, though minor, source of variability for the timing of spikes originates from an uncertainty of the state of the integrator at the beginning of a trial.

\begin{figure}
  \begin{center}
    \includegraphics[width=\textwidth]{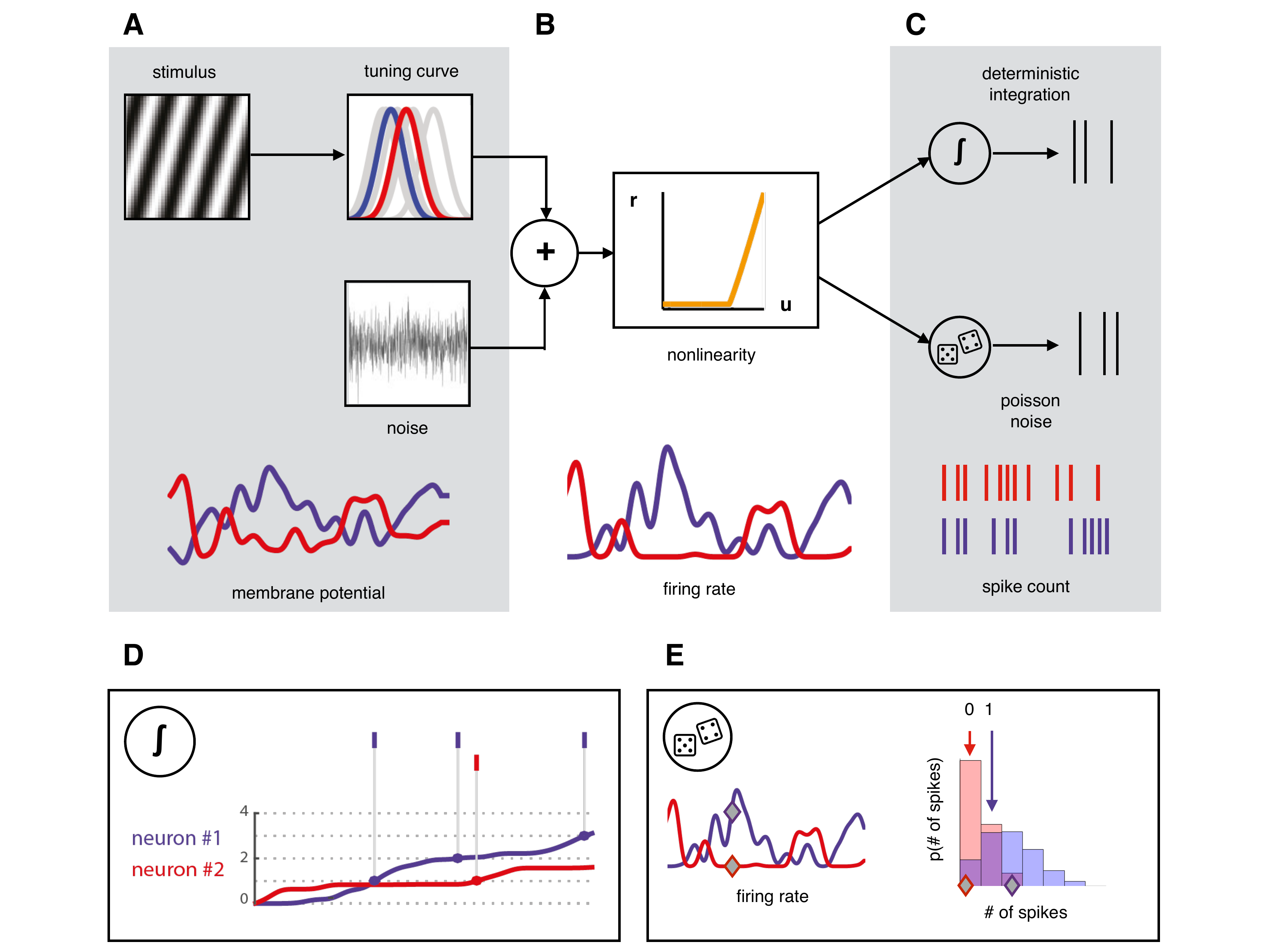}
  \end{center}
  \caption{\textbf{Schematic illustration of the RG and DSP models.} \textbf{\textit{A}}, Membrane potential responses are characterized by a mean activation and a stochastic component which varies over time. The mean is assumed to be stimulus-dependent, and is determined by the orientation tuning curve. Importantly, the stochastic component is not independent but is correlated among neurons. \textbf{\textit{B}}, Firing rates are calculated from membrane potentials by transforming them using the firing rate nonlinearity. \textbf{\textit{C}}, The way spike counts are obtained from the firing rate is determined by the spiking model. \textbf{\textit{D}}, According to the Rectified Gaussian (RG) spiking model, spikes are generated by integrating the firing rate, and deterministically registering a spike every time the integrated rate crosses an integer value. \textbf{\textit{E}}, In the Doubly Stochastic (DSP) spiking model, spike counts are stochastically generated: the time varying firing rate (normalized by the time window being considered, diamonds, left panel) determines the mean of the Poisson distribution (diamonds, right panel), which assigns probabilities to the number of spikes to be generated in the time window (right panel).}
  \label{fig:cartoon}
\end{figure}

A cartoon depicting the procedure of generating firing rates from Gaussian membrane potentials, and then spike counts using the two models, is summarized in Fig.~\ref{fig:cartoon}.

\subsection{Simulation of population measurements}

In order to compare the predictions of the RG and DSP models with experimental recordings of activity of a population of neurons, population models were constructed and spike response statistics of the two models were contrasted with experimental data. The primary goal of these comparisons was to test predictions of the two models on the population level, where summary statistics are less susceptible to sampling noise, and their measurements do not require the precise control of receptive field contents. We simulated full-field gratings as stimuli in the simulated experimental paradigm. 

Determining the membrane potential statistics for the population of $N$ neurons requires the specification of their means, variances and correlations, as specified in Eq.~\ref{eq:mp_gauss}. The mean membrane potential response $\mu^n$ for cell $n$ is determined by stimulus orientation through the tuning curve that is characteristic to the particular neuron. The stimulus orientation at which the  peak of the tuning curve is located corresponds to the preferred orientation of the neuron and preferred orientations are sampled independently for each neuron from a uniform distribution. The height of the tuning curve is varying from neuron to neuron and is sampled from a Gaussian distribution with a standard deviation of 0.1. Widths of the tuning curve is fixed at  $SD_{TC} = 0.2\pi$. The time varying stochastic component of the membrane potential response is coming from a Gaussian distribution which is characterized by its variance. The level of variance was inhomogeneous across the population and was set randomly by sampling  an inverse gamma distribution. Parameters of the inverse gamma distribution are the shape and scale parameters, with values three and four, respectively, chosen to reproduce the scale of spike count Fano factors observed in experiments. Membrane potential correlation matrices are generated algorithmically by specifying the width of the distribution with a scalar parameter \citep{Lewandowski:2009gd}. Distribution of membrane potential correlations were tuned such that the distributions of spike count correlations were matched for the RG and DSP models. 

Simulation of changes on stimulus orientation is straightforward since tuning curves define the changes in membrane potential mean, while other aspects of the statistics are assumed to be unchanged. This choice is motivated by studies on membrane potential variance, which demonstrated that variance is relatively intact by changes in stimulus orientation \citep{Finn:2007hc}. The question of the orientation dependence of membrane potential correlations is still open but we took a conservative approach by assuming orientation independence at the level of membrane potential correlations. Modulation of stimulus contrast was simulated by adjusting both the gain of the tuning curve and the level of variance. Lowered contrast caused the mean membrane potential response to scale down to a level half of that at high contrast, while the variance increased to 1.4 times the original level. 

When the membrane potential statistics of the cell populations are fully defined, we simulate membrane potential responses by taking membrane potential samples from the population. A single trial was 500 ms long, the number of trials was 1000. The samples are then transformed to instantaneous firing rates by the rectifier nonlinearity (Eq.~\ref{eq:firing_rate}), and then spike counts are obtained by using the DSP and RG spiking models.

\subsection{Electrophysiological data}

To test the predictions of the models, we used publicly available data recorded in the labs of Matthias Bethge and Andreas Tolias \citep{Ecker:2010dn}. Detailed description of the recording settings are available at the original publication. Briefly, unit recordings were obtained by extracellular electrode arrays from the primary visual cortex of awake monkeys. Stimuli consisted of static and moving full-field gratings. We constrained our analysis to static gratings because the static grating data set featured multiple contrast levels besides eight grating orientations. We used 400 ms segments extracted from the evoked activity period of the trials in which the spike counts were calculated. 

In order to reliably estimate pairwise correlations, we needed to exclude some of the recordings. We only considered pairs in which both units had an average firing rate over 0.1 Hz to avoid biased correlation estimates due to the insufficient number of spiking events. Pilot analyses (data not shown) have demonstrated that low number of stimulus repetitions can lead to highly inconsistent estimates of the spike count correlations, therefore we only included recording sessions that consisted of at least 39 repetitions.Thus, we included five sessions in the analysis, with repetition numbers (39, 40, 85, 72, 39). The filtering criteria for firing rates and trial numbers allowed us to use 41 units from the recordings.

When comparing spike count correlations between subpopulations observing a stimulus with a preferred or a non-preferred orientation, we defined orientations as preferred when the firing rate of a unit averaged over the trials using the orientation was higher than the average firing rate over all trials. This binary classification scheme helped us to avoid errors in orientation preference estimation, and in the same time lead to more reliably estimated of correlations in the non-preferred condition, as including only an orientation perpendicular to the most preferred one would have produced very low firing rates. Preferred-orientation correlations were calculated between pairs both observing a preferred stimulus.

\subsection{Analysis of neural responses}

We characterize the distributions of spiking responses of neurons up to second-order statistics, similarly to the descriptions of membrane potentials. However, due to specific properties of spike trains, the applied measures are slightly different. Spike count responses are characterized by variances that grow linearly with spike count means. Therefore, we are interested in changes in the variance that are independent of changes in the mean. By using Fano factor,we can control for this effect and can obtain a trial-by-trial measure of response variability for neuron $n$:

\begin{equation}\label{eq:ff}
\mathrm{FF} \left[ s^n_t \right]_t = \frac{\mathrm{Var} \left[ s^n_t \right]_t}{\mathrm{E} \left[ s^n_t \right]_t}
\end{equation}

The choice is also supported by the fact that systematic changes in the membrane potential variance are similarly observed in the spike count Fano factor, as described by \citep{Churchland:2010he}.

The pairwise co-activation of a pair of neurons given their spike trains is fully characterised by the cross-correlogram of their spike counts. While correlations may occur at different time scales, it is a typical choice in experiments to use the correlation of spike counts over entire trials. Doing so has the advantage of taking interactions with different delays into account similarly, by sacrificing the finer temporal structure of co-activations \citep{Smith:2008gv}. In order to account for irregularities in the firing rates and individual variances of experimentally recorded spike trains, correlations are calculated between z-scored spike counts, defined as follows:

\begin{equation}\label{eq:sc_corr}
\varrho^{ij} = \mathrm{Corr} \left[\frac{s^i_t - \mathrm{E} \left[s^i_t \right]_t }{\mathrm{SD} \left[s^i_t \right]_t}, \frac{s^j_t - \mathrm{E} \left[s^j_t \right]_t }{\mathrm{SD} \left[s^j_t \right]_t} \right]_t
\end{equation}

Aiming for population-level characterisation of responses puts constraints on what kind of measurements are applicable for comparison. As anesthesia is known to introduce significant biases in neuronal response correlations \citep{Ecker:2014cl}, we sought to test model predictions against  data  recorded from awake animals.

\section{Results}

Extensive data on the linear relationship between spike count mean and spike count variance \citep{Softky:1993uj, Britten:1993wv, Tolhurst:1983wa} motivated a model of spiking activity that assumes a Poisson process at spike generation. In this model, the Poisson-like single-cell spiking statistics can be attributed to private noise, a form of noise that is not shared by neurons. As a consequence, it does not necessarily contain any other forms of variability. Such a setting constrains the model to a Fano factor that is equal to one (when signal variance related to changing stimuli is ignored) and no variability that is correlated across neurons. V1 simple cells, however, can express correlated activity \citep{Ecker:2010dn} and are characterized by Fano factors that both deviate from one and can change with changing stimulus attributes. In order to accommodate these effects, we extended the simple Poisson model: the Doubly Stochastic Poisson (DSP) model assumes that the membrane potential has a stochastic component (Fig.~\ref{fig:cartoon}E). The membrane potential was sampled from a multivariate normal distribution, and consecutive samples were assumed to be independent across 20 ms time bins. Independence of samples is a simplifying assumption that is motivated by the fast-decaying autocorrelation function of V1 neurons \citep{Azouz:1999wj}. The membrane potential of a model simple cell was transformed by a nonlinearity to obtain a firing rate \citep{Carandini:2000ud} and spiking activity was obtained by the Poisson process, which ensured an expected value for the number of spikes proportional to the rate. The threshold-power-law firing rate nonlinearity ensured the correct mapping between mean membrane potential and firing rate \citep{Carandini:2004ee}. 

In the alternative Rectified Gaussian (RG) model, there was a single source of stochasticity which was a stochastic process at the level of membrane potential. Again, the distribution at any given time bin was a multivariate normal distribution and time bins were independent (Fig.~\ref{fig:cartoon}A). Membrane potentials were mapped through the same nonlinearity as in the DSP model which ensured a similar evolution of firing rate with increased mean membrane potential (Fig.~\ref{fig:cartoon}B). Spikes were obtained by a process that bears as little stochasticity as possible: firing rate was integrated over time and spikes were generated when the integral crossed integer values (Fig.~\ref{fig:cartoon}D). While this model does not rely on a Poisson process to ensure the scaling of spike count variance with the mean, it has been demonstrated to account for the linear relationship between spike count mean and spike count variance solely as a result of the interaction of the subthreshold variability and the firing rate nonlinearity \citep{Carandini:2004ee}. 

\begin{figure}
  \begin{center}
    \includegraphics[width=\textwidth]{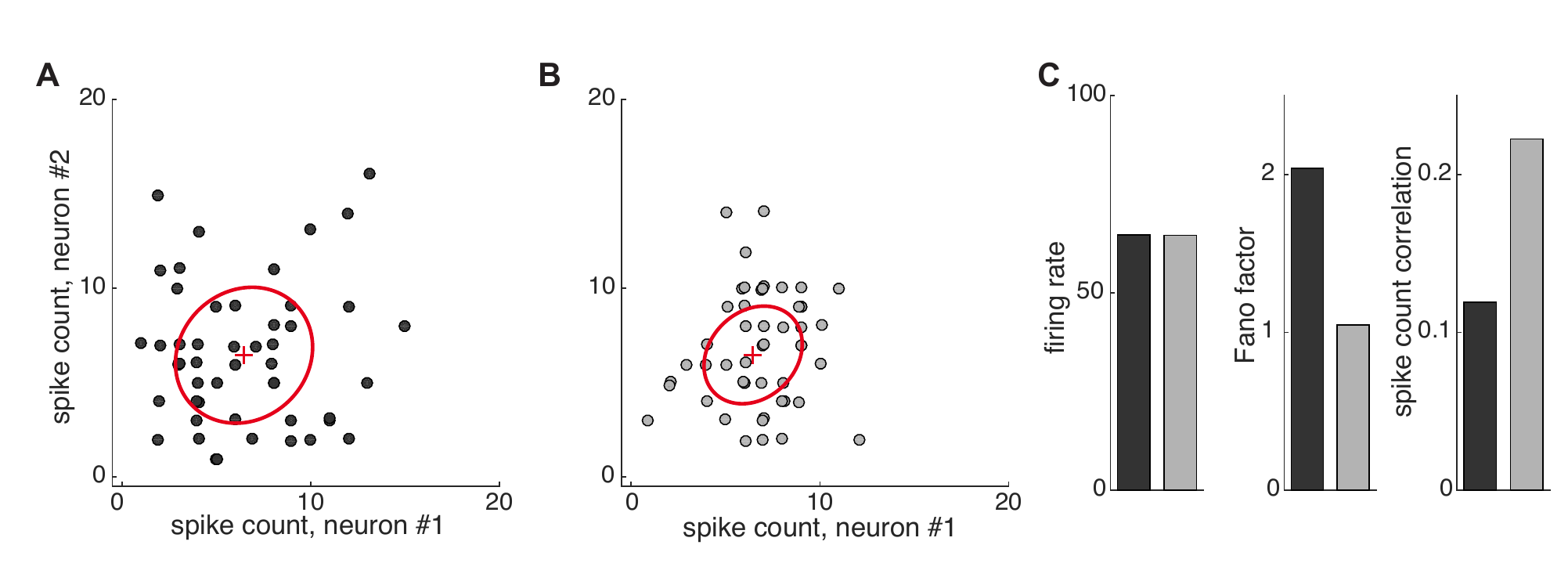}
  \end{center}
  \caption{\textbf{Spike response statistics for the Doubly-stochastic Poisson (DSP) and Rectified Gaussian (RG) models.} \textbf{\textit{A-B}}, Spike count distribution for a pair of model neurons with the same membrane potential statistics (mean, variance, correlation) for DSP (\textbf{\textit{A}}) and RG models (\textbf{\textit{B}}) in a simulated 100-ms time window. Membrane potential means and variances were identical for the two neurons and membrane potential correlation was set to 0.25. Circles indicate spike counts for individual trials. Cross shows the across-trial mean while ellipse represents the across-trial covariance ellipse of the joint spike count distribution. Small jitter was added to spike counts for illustration purposes.  \textbf{\textit{C}}, While the spiking models are consistent in predicting equal mean spike count responses for both DSP (dark bars) and RG models (light bars) at matching membrane potential distributions, Fano factors and spike count correlations show characteristic differences. Note that spike count correlations are systematically lower than membrane potential correlations at both models.  }
  \label{fig:mpmatch}
\end{figure}

The expressive power of the two models is similar, which is also confirmed by the equal number of parameters characterizing the two models. These parameters were established based on previously published data (see Materials and Methods) \citep{Carandini:2004ee}. There are important differences, however, in the statistics of spiking activities the two models predict (Fig.~\ref{fig:mpmatch}A,B). Assuming identical distributions for membrane potentials for a pair of model neurons, the firing rate nonlinearity and consistent spike generation processes ensure similar mean spike counts (Fig.~\ref{fig:mpmatch}C). The variances, however, differ for the two models: while the variance of responses of RG model neurons is dominantly determined by the appropriately scaled variance of the membrane potentials, the variance of the DSP model is the sum of the membrane potential variance and a term coming from the Poisson stochasticity. As a result, the spike count variance, as well as the Fano factor, of the DSP model exceeds that of the RG model (Fig.~\ref{fig:mpmatch}C). The correlation measured from the spike count distribution can differ from the membrane potential correlations as a result of multiple factors \citep{Cohen:2011eh, Ecker:2010dn}. Most importantly, the firing rate nonlinearity can truncate the subthreshold part of the membrane potential distribution causing a decrease in spike count correlations relative to the membrane potential correlations, which is evident for the RG model (Fig.~\ref{fig:mpmatch}B).  Another important consequence of the excess private variability introduced by the Poisson spike generation process is a further drop in the spike count correlation (Fig.~\ref{fig:mpmatch}C). A simple intuition for this effect can be obtained by considering that the covariance matrix of the spike count correlation is the sum of the membrane potential covariance and the covariance of the spike generation. The latter, however, is a diagonal matrix, since this source of noise is independent among neurons, therefore it only increases the diagonal elements of the resulting covariance matrix, and since correlation is the ratio of the covariance and the geometric mean of the variances, there is an overall decrease in correlations. 

In the coming sections we analyze the consequences of these differences on simplified model of a pair of neurons before moving to the analysis of the responses statistics of populations of neurons. When using the simplified model we do not simulate the tuning curve-mediated changes in membrane potentials, rather we directly investigate the effects of changes in membrane potential statistics. These analyses provide predictions on changes in spike count statistics expected in response to changes in stimulus orientation and contrast.

\subsection{Matching spike response statistics}

In order to be able to contrast the effects of stimulus change on response statistics of the competing models, we first establish a method for matching the spiking statistics of the DSP and RG models in a pair of neurons. Since equal membrane potential statistics lead to different spiking statistics, it is clear that either membrane potential statistics or parameters of the firing rate nonlinearity need to be adjusted to have matching firing rates, Fano factors and spike count correlations. In order to keep our arguments simple, we keep the firing rate nonlinearity unchanged. In fact, the scale of membrane potential (which is determined together by the distance of the membrane potential from the firing rate threshold and the variance of the membrane potential) and the scale of firing rate nonlinearity (parameter $k$) can be altered largely interchangeably (see Fig.~\ref{fig:corrdep}), therefore the argument can be translated into changes in the scale parameter of the firing rate. 

\begin{figure}
  \begin{center}
    \includegraphics[width=\textwidth]{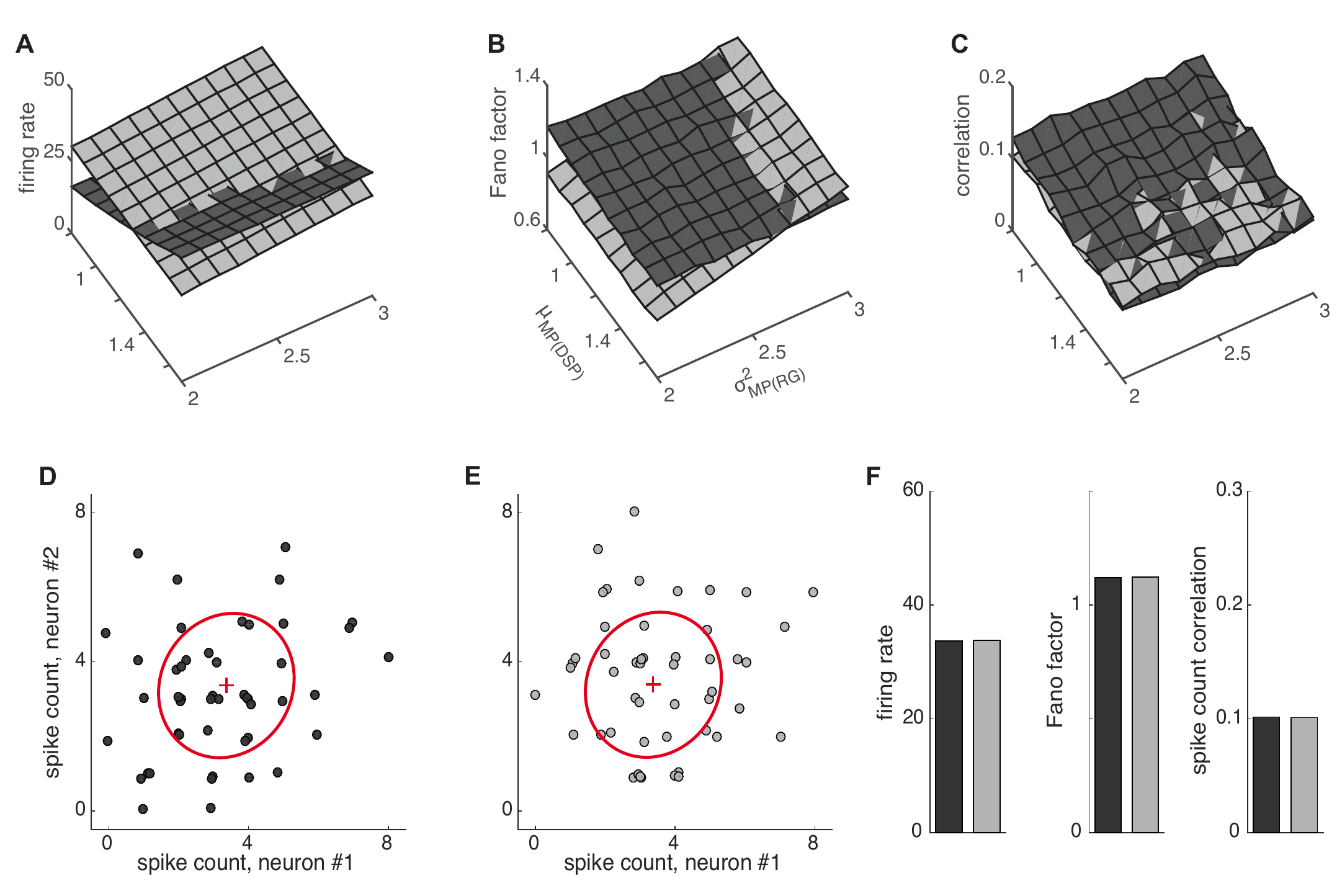}
  \end{center}
  \caption{\textbf{Matching the output statistics of  DSP and RG models.} \textbf{\textit{A-C}}, Spike count (\textbf{\textit{A}}), Fano factor (\textbf{\textit{B}}) and spike count correlation (\textbf{\textit{C}}) of the neurons as a function of membrane potential parameters for the DSP (dark plane) and RG model (light plane). Simulations show responses in a 100 ms time window. Intersections of the plane indicate equal statistical measure for the two models which then help to determine the spike count statistics. Parameters explored for matching spiking statistics were  the mean membrane potential of the DSP model and membrane potential variance of the RG model. \textbf{\textit{D-F}}, By tuning membrane potential statistics for the DSP (\textbf{\textit{D}}) and RG (\textbf{\textit{E}}) models individually, the spike count statistics of the models can be matched up to the mean, variability (as measured by the Fano factor) and correlation. Spike count distribution representation is the same as that on Fig.~\ref{fig:mpmatch}. Matched means. \textbf{\textit{F}}, spiking statistics for the matched models. Colours are identical to panels \textbf{\textit{A-C}}.}
  \label{fig:statmatch}
\end{figure}

By exploring membrane potential parameters for the two models, we can obtain a parameter setting where firing rates (Fig.~\ref{fig:statmatch}A) and Fano factors (Fig.~\ref{fig:statmatch}B) are matched. Both of these criteria constrain the parameter sets up to a linear combination of the tested parameters, therefore the intersection of the lines allows matched firing rate and Fano factor. Difference between the spike count correlations along the explored parameter range is relatively untouched (Fig.~\ref{fig:statmatch}C), and can be adjusted by tuning membrane potential correlations. The resulting statistics-matched models (Fig.~\ref{fig:statmatch}D-F) have markedly different membrane potential variances (0.2 $\mathrm{mV^2}$ and 2.75 $\mathrm{mV^2}$ for the DSP and RG models, respectively) and different membrane potential correlations (0.95 and 0.13 for the DSP and RG models, respectively). Because of the extra variance and the decorrelation effect of the DSP model, these differences are expected and show that the Poisson process introduces a private variability which can easily wash out membrane potential correlations. While the membrane potential correlation level required in the DSP model seems to be extreme, it only serves the purpose of matching the spiking statistics. At lower firing rates the excess variance added to the membrane potential covariance would be lower and  therefore  statistics matching would require considerably lower membrane potential correlations. 

\subsection{Dependence of spiking statistics on the membrane potential mean}

The membrane potential mean in V1 simple cells is sensitive to stimulus orientation. In order to understand the effects incurred by orientation change on spiking statistics, we need to separate the effects on different aspects of the response statistics. Changes in membrane potential mean have obvious effects on the firing rate. Effects of mean membrane potential on other characteristics of the response statistics are less straightforward. In order to be able to see how the membrane potential mean changes affect Fano factors and spike count correlations in DSP and RG models, we aim to see changes in these measures independent of changes in firing rates. 

\begin{figure}
  \begin{center}
    \includegraphics[width=\textwidth]{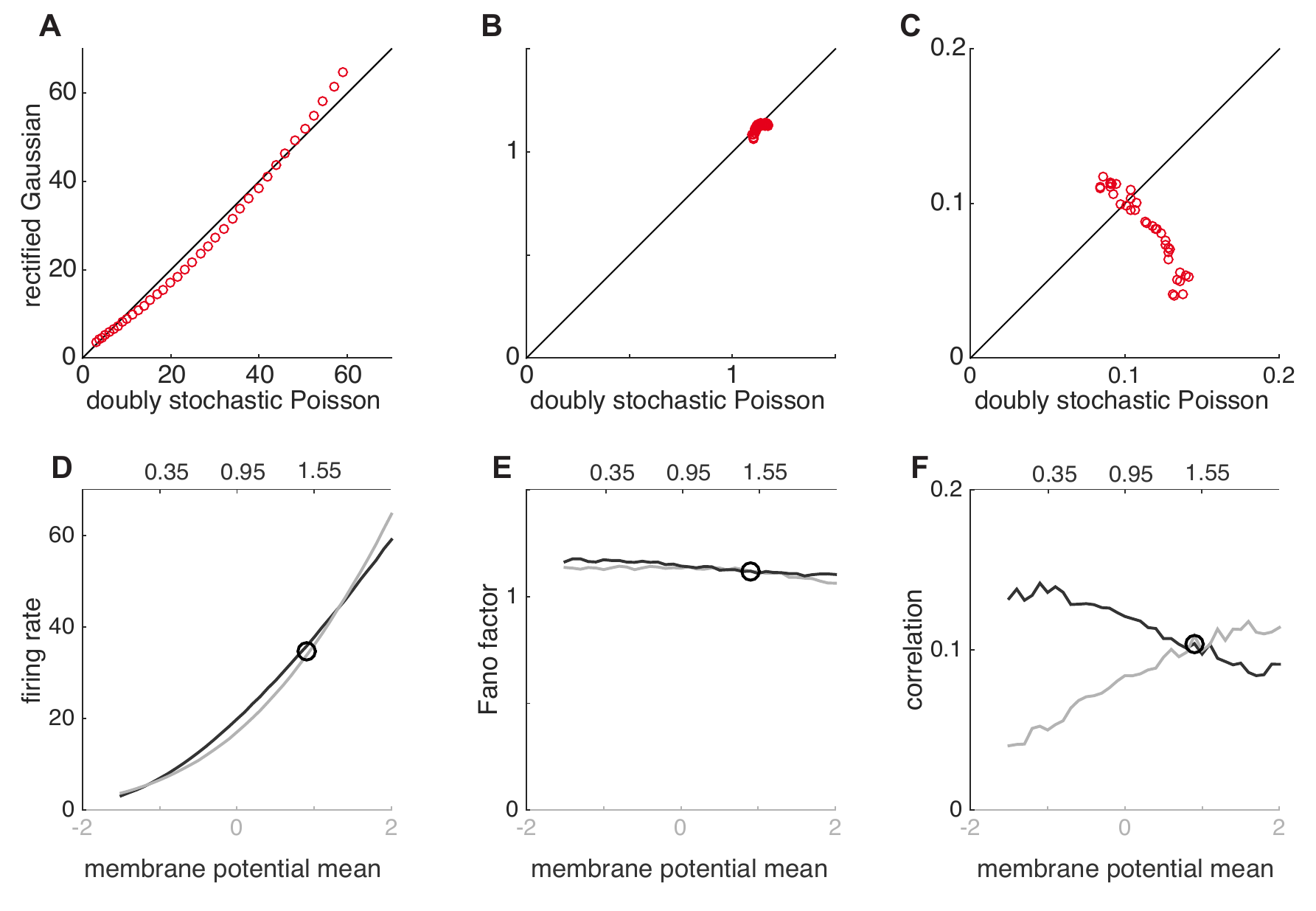}
  \end{center}
  \caption{\textbf{Dependence of spike count statistics on the membrane potential mean.} \textbf{\textit{A-C}}, Predictions of the two spiking models directly contrasted against each other for firing rate (\textbf{\textit{A}}), Fano factor (\textbf{\textit{B}}), and spike count correlation (\textbf{\textit{C}}) upon changes in membrane potential mean. Matched-statistics DSP and RG are tested with changing membrane potential means but constant membrane potential variances. \textbf{\textit{A}}, With appropriate linear scaling of the membrane potentials, approximately equal firing rates can be achieved across a range of membrane potentials resulting in a wide range of firing rates (0 to 60 Hz). \textbf{\textit{B}}, Fano factors are close to equal and invariant across the whole range of membrane potentials. \textbf{\textit{C}}, While firing rates and Fano factors cannot explicitly differentiate between the DSP and RG models, the correlation shows systematic differences. \textbf{\textit{D-F}}, Same as \textbf{\textit{A-C}} but respective statistics for the DSP (dark lines) and RG (light lines) models as a function of the membrane potentials. Note that in order to match the spiking statistics, the membrane potential statistics (including the mean, variance, and correlation of neurons) differ in the two models (see different horizontal axes at the bottom and the top of panels). \textbf{\textit{D}}, Firing rates from the two different model neurons show characteristic nonlinearity of firing rates in the two models. \textbf{\textit{E}}, Fano factors in both models show only minimal dependence on membrane potential (and thus on firing rate). \textbf{\textit{F}}, Spike count correlations show sensitivity to changes in the membrane potential: while pairwise correlation in the DSP model decreases with increasing membrane potential levels, it increases in the RG model. Except for the changing mean membrane potential, parameters of the models are the same as those used on Fig.~\ref{fig:statmatch} (matched mean is denoted by black circle).}
  \label{fig:orientchange}
\end{figure}

Membrane potential statistics are different in the statistics-matched DSP and RG models. Therefore, we first establish rates of change for the membrane potentials in the two models, which guarantee that even upon deviating from the statistics-matched levels of membrane potentials, the firing rates change at a similar rate (Fig.~\ref{fig:orientchange}A,D). Using this firing rate-matched scenario, we can directly contrast mean-related changes in Fano factors (Fig.~\ref{fig:orientchange}B,E) and spike count correlations (Fig.~\ref{fig:orientchange}C,F). While Fano factors are approximately equal (Fig.~\ref{fig:orientchange}B) across the range of membrane potentials for the DSP and RG models, spike count correlations deviate for the two models (Fig.~\ref{fig:orientchange}C). Independence of Fano factors from membrane potentials (Fig.~\ref{fig:orientchange}E), and from firing rates as well, confirms the original results of the RG model \citep{Carandini:2004ee} and is expected for the DSP model. Membrane potential dependence of spike count correlations (Fig.~\ref{fig:orientchange}F) reveals a decreasing tendency for the DSP and an increasing tendency of the RG with increasing membrane potential levels. Growing firing rate resulting from increased mean membrane potential has a differential effect in the two models. In the DSP, increased firing rate incurs increased private variability which suppresses the contribution of the membrane potential covariance to the total covariance and therefore spike count correlation diminishes. In the RG model, however, a different mechanism dominates: increased mean activation results in a higher proportion of the membrane potential covariance to be above firing threshold, and consequently, smaller truncation of this distribution boosts the magnitude of the measured correlation \citep{deLaRocha:2007go}. Taken together, under controlled change in mean activity, spiking variability is relatively insensitive to the choice of DSP or RG models. Analysis of spike count correlations, however, provide opposing predictions in the case of DSP and RG models for manipulations that affect mean responses.

\begin{figure}
  \begin{center}
    \includegraphics[width=\textwidth]{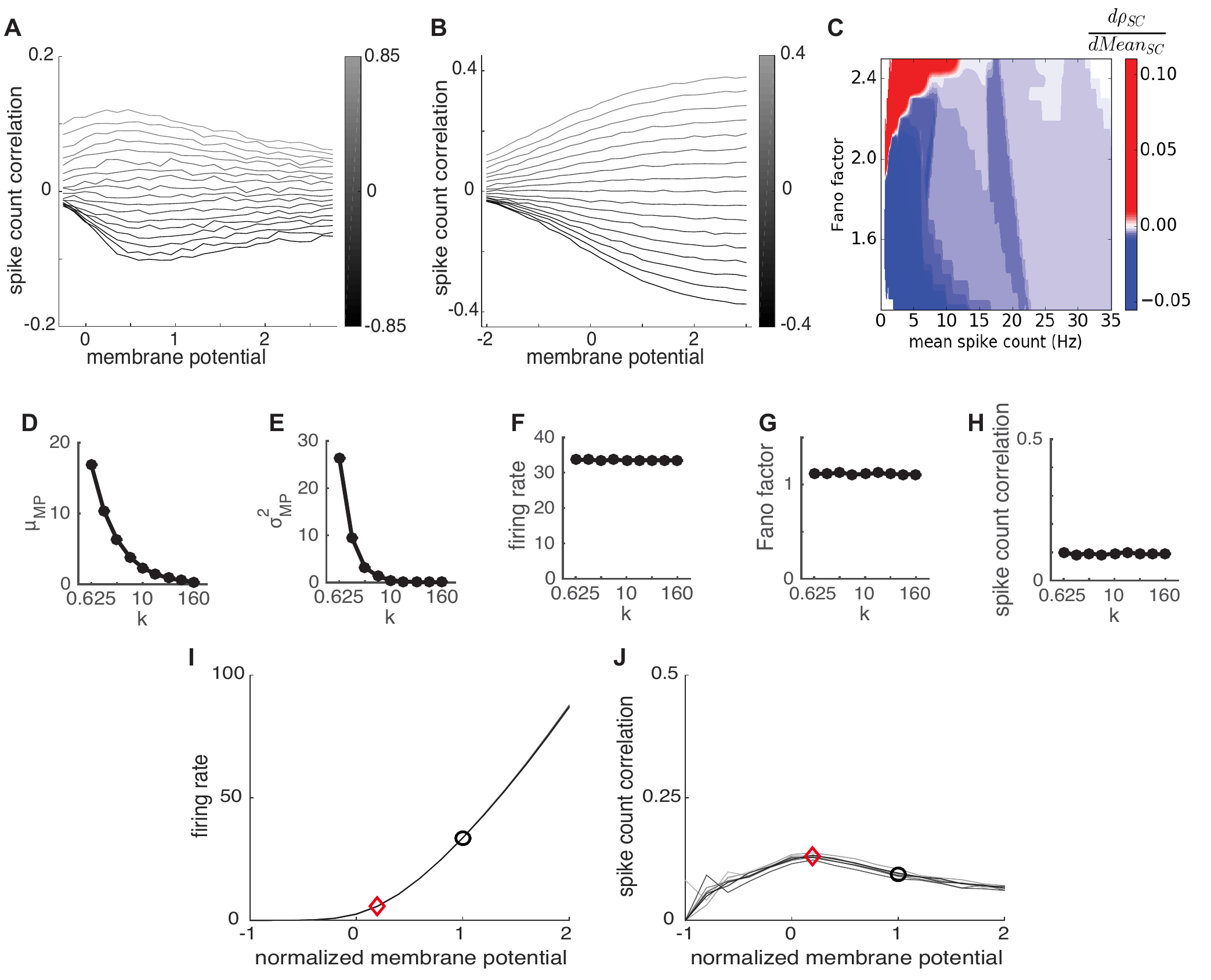}
  \end{center}
  \caption{\textbf{Relationship between spike count correlations and membrane potential mean in the two spiking models.} Evolution of spike count correlations with changing membrane potentials at different membrane correlation levels in the matched statistics DSP (\textbf{\textit{A}}) and RG (\textbf{\textit{B}}) models. \textbf{\textit{A}}, In the DSP model, after a brief increase of the correlation magnitude towards the level of membrane potential correlation, the tendency reverses and a decline towards zero spike count correlation takes place. Range of membrane potential correlations tested spans from -0.95 to 0.95. \textbf{\textit{B}}, The RG model is characterized by a steadily increasing magnitude of spike count correlation, saturating at the correlation of the membrane potentials. Range of membrane potential correlations tested spans from -0.4 to 0.4. \textbf{\textit{C}}, Exploration of spike count statistics in the DSP model to assess the regimes in which correlation magnitude increases with the firing rate. Positive slope regimes (red hues) are constrained to low firing rates and high Fano factors, with the white separating blue and red regimes denoting the peak of the curve in \textbf{\textit{A}}. \textbf{\textit{D-J}}, Analysis of the dependence of spike count statistics on the base rate. Membrane potential statistics (\textbf{\textit{D-E}}) are tuned for a range of values of the base rate parameter $k$ in order to produce similar spiking statistics (\textbf{\textit{F-H}}). \textbf{\textit{I}}, Membrane potential dependence of firing rates is identical for the different gain, $k$ (lines are overlapping). Membrane potentials are normalized to the values established on \textbf{\textit{D}} (black circle), diamond denotes the maximum on \textbf{\textit{J}}. \textbf{\textit{J}}, Membrane potential dependence of correlations is similar across different levels of the gain, $k$ (grey lines). }
  \label{fig:corrdep}
\end{figure}

Differential effects of the membrane potential mean on spike count correlations in the two models highlight an opportunity to distinguish between the models upon changes in stimulus orientation. A detailed exploration of the evolution of spike count correlation with changing membrane potential mean can reveal the generality of the effect seen on Fig.~\ref{fig:orientchange}F. We tested this question by assessing spike count correlations at different levels of membrane potential correlations (Fig.~\ref{fig:corrdep}). Analysis of the RG model reveals a monotonic rise of the magnitude of spike count correlations from zero towards the level of the membrane potential correlation as membrane potential mean increases (Fig.~\ref{fig:corrdep}B). In the DSP model, the effect of a less truncated membrane potential joint distribution, when a larger proportion of the distribution gets above threshold, is shown at a low membrane potential regime (Fig.~\ref{fig:corrdep}A): at low levels of activations, a rise similar to the RG model can be observed. This range, however, is severely limited (Fig.~\ref{fig:corrdep}C) and can be observed at moderate firing rates. Beyond that point, a steady decline of spike count correlation takes place which converges to zero (Fig.~\ref{fig:corrdep}A,C). The biphasic profile of membrane potential dependence of spike count correlation raises the possibility that an increased gain in the firing rate nonlinearity can simply scale the firing rate profile. Thus, the regime where the spike count correlation is positively correlated with membrane potential mean could possibly overcome the limited range shown on Fig.~\ref{fig:corrdep}A and could reach higher firing rates. We tested this question by scaling the firing rate gain together with inverse scaling of the membrane potential (Fig.~\ref{fig:corrdep}D). In order to keep not only the firing rate but also the Fano factor constant, the variance of the membrane potential was also scaled together with the mean and rate gain parameters (Fig.~\ref{fig:corrdep}E). In the resulting setting we could test a wide range of the gain parameter $k$, while keeping the mean, the Fano factor, and the correlation of spiking responses constant (Fig. F-H). The firing rate profile was identical for the different parameter settings (Fig.~\ref{fig:corrdep}I). Importantly, the evolution of correlations was very close at different settings of the gain parameter $k$, with no visible shift in the membrane potential (or alternatively firing rate) value maximizing the correlation  (Fig.~\ref{fig:corrdep}J). This analysis demonstrates that the regime where increasing correlations are present with increasing firing rates are constrained to low firing rate levels and high Fano factors.

\subsection{Dependence of spiking statistics on membrane potential variance}

Stimulus contrast was demonstrated to have a combined effect on membrane potential mean and variance \citep{Finn:2007hc}: while the mean of the membrane potential response shrinks as contrast goes to zero, membrane potential variance grows. Thus, changes in membrane potential response variance have relevant consequences on the spiking statistics. In order to get insights into the effects of joint changes in membrane potential mean and membrane potential variance, we first explored these characteristics separately and then turned to the combined effects of parallel changes. 

\begin{figure}
  \begin{center}
    \includegraphics[width=\textwidth]{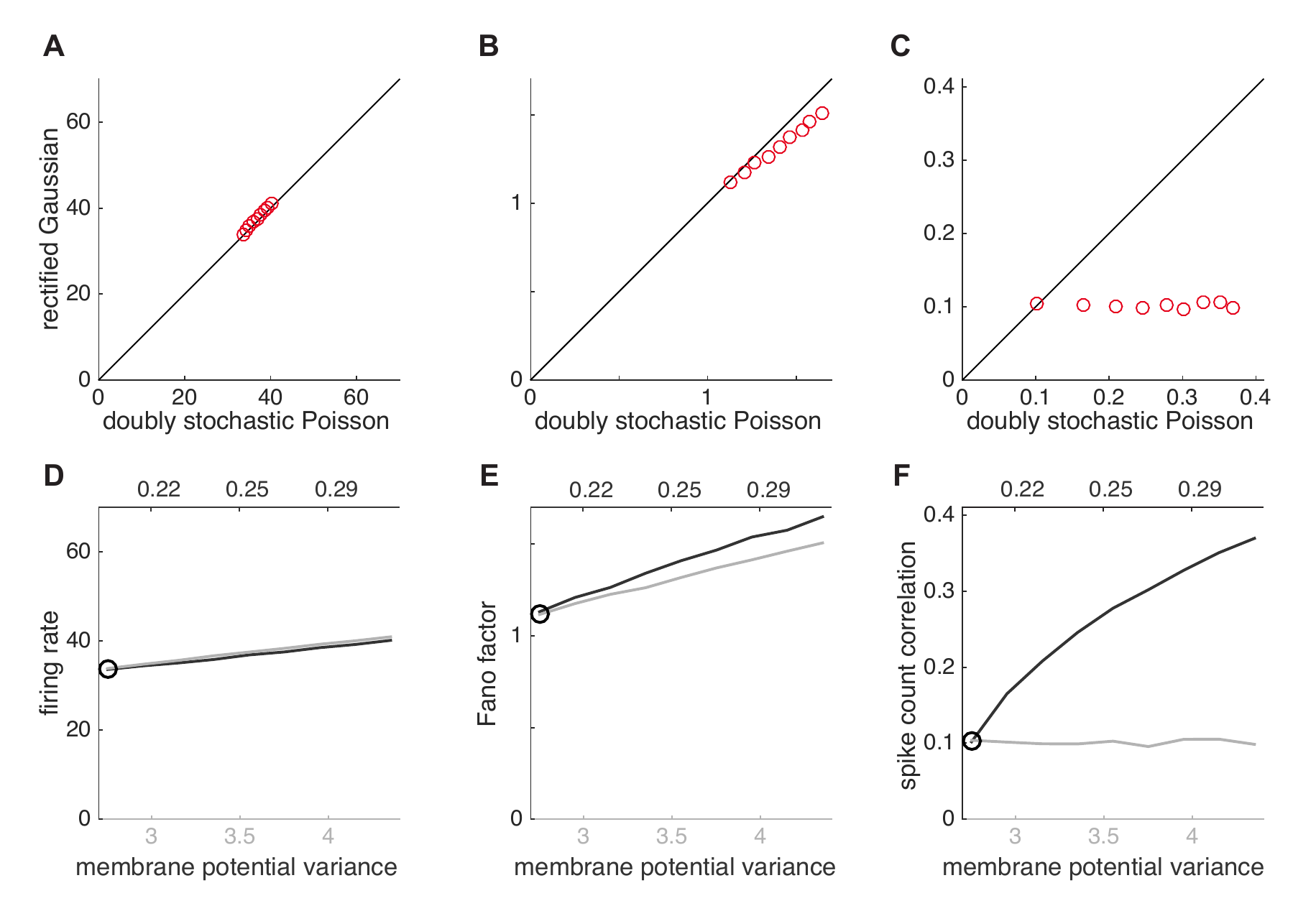}
  \end{center}
  \caption{\textbf{Dependence of spike count statistics on the membrane potential variance.} Spiking statistics of the two spiking models directly contrasted against each other for the two models upon changes in membrane potential variance. Matched-statistics DSP and RG are tested with changing membrane potential variances but constant membrane potential means. Firing rates (\textbf{\textit{A}}) and Fano factors (\textbf{\textit{B}}) are approximately equal  in both models, but spike count correlations (\textbf{\textit{C}}) show systematic differences. \textbf{\textit{D-F}}, Plotting the statistical measures against membrane potential variance reveals that both firing rates (\textbf{\textit{D}}) and Fano factors (\textbf{\textit{E}}) show a tendency to increase with increased membrane potential variance. Dark lines: DSP model; light lines: GR model. Note that in order to match the spiking statistics, the membrane potential statistics (including the mean, variance, and correlation of neurons) differ in the two models (see different horizontal axes at the bottom and the top of panels). While spike count correlations are invariant across the range of variances for the RG model, the poisson model shows a steady increase with increasing membrane potential variance (\textbf{\textit{F}}). Except for the changing mean membrane potential, parameters of the models are the same as those used on Fig.~\ref{fig:statmatch} (matched variance is denoted by black circle).}
  \label{fig:variancechange}
\end{figure}

After the mean-dependent changes discussed in the previous section, we set out to analyze the membrane potential variance-dependence of spiking statistics (Fig.~\ref{fig:variancechange}). Similarly to the protocol followed at testing the effects of membrane potential mean, we started from the matched-statistics DSP and RG models, and set the range of variance scaling such that the resulting firing rate changes in the two models are approximately equal (Fig.~\ref{fig:variancechange}A,D). Again, this mean firing rate-matched approach ensures that changes seen in the Fano factors and spike count correlations are not related to differences in firing rates. Contrasting the Fano factors at firing rate-matched settings of the DSP and RG models revealed similar tendencies but slightly differing values for the Fano factors (Fig.~\ref{fig:variancechange}B). Increased membrane potential variance translated into increased Fano factors in both of the spiking models, but the DSP model was characterized by systematically larger Fano factors at higher membrane  potential variances (Fig.~\ref{fig:variancechange}E). This difference is due to the excess variance of the DSP model coming from the increased spike count variance of the Poisson stochasticity at higher mean spike counts. Changing membrane potential variance showed conflicting effects in the two models on spike count correlations (Fig.~\ref{fig:variancechange}C). While the spike count correlation in the RG model was relatively insensitive to changes in membrane potential variance and thus firing rate, the DSP model was shown to exhibit increased spike count correlation with increased membrane potential variance (Fig.~\ref{fig:variancechange}F). This increase can be easily understood by recognizing that a scaled membrane potential covariance results in a larger relative contribution of the membrane potential covariance to the total covariance, thus the spike count correlation will be more dependent on the correlated membrane potential stochasticity than uncorrelated spiking stochasticity. 

\begin{figure}
  \begin{center}
    \includegraphics[width=\textwidth]{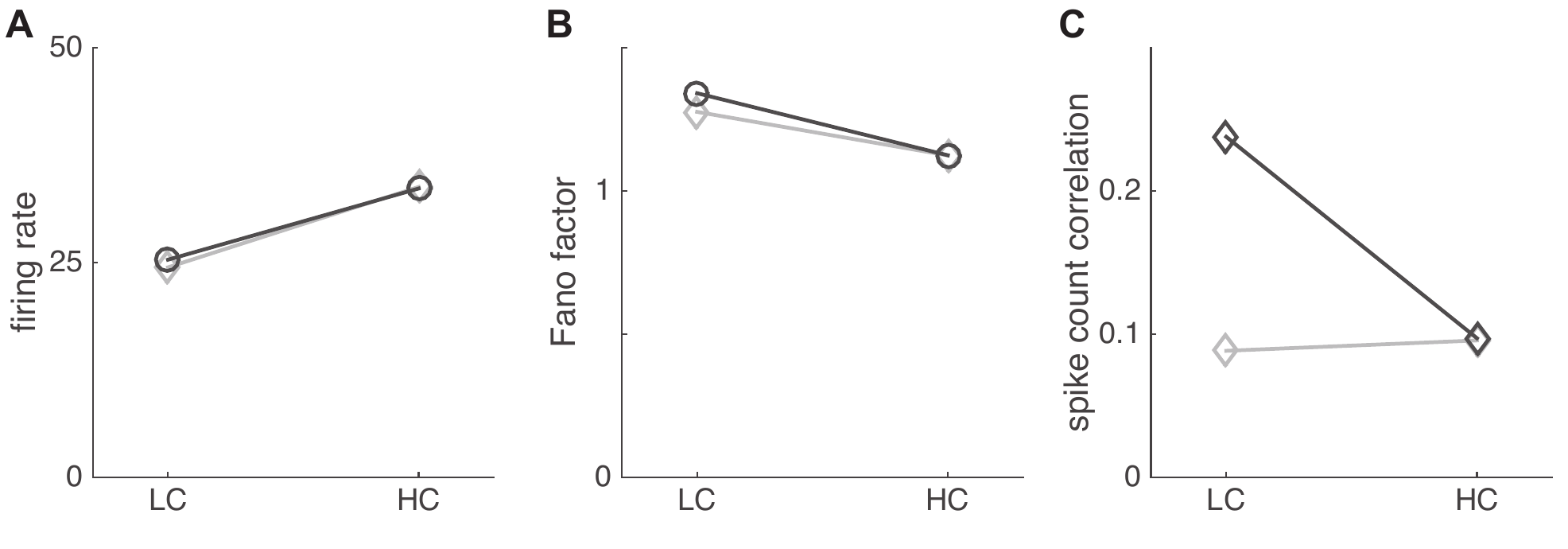}
  \end{center}
  \caption{\textbf{Contrast dependence of spiking statistics.} Spike counts, Fano factors and spike count correlations in the two models (dark and light lines for the DSP and RG models, respectively) with concomitant changes in membrane potential mean and variance stimulating a switch from high contrast (HC) to low contrast (LC): decreased mean was followed by increased variance. Parameters for high and low contrast levels were set such that firing rate changes (\textbf{\textit{A}}) and Fano factor changes (\textbf{\textit{B}}) approximated the changes expected in V1 recordings. Under such conditions the spike count correlation is expected to decrease for the DSP model but increase for the RG model when contrast is increased (\textbf{\textit{C}}). }
  \label{fig:lchc}
\end{figure}

Contrast change incurs concomitant changes in membrane potential mean and variance. Based on the previous analyses we can conclude how these parallel changes interact when the spiking statistics are considered in response to a stimulus at lower contrast. In terms of firing rate, decreased mean membrane potential and increased variance due to reduced contrast have opposing effects, but the effect of decreased membrane potential dominates patterns in firing rate (Fig.~\ref{fig:lchc}A). In terms of Fano factor, it is the change in variance that causes increased Fano factors in both models (Fig.~\ref{fig:lchc}B). In terms of spike count correlations, the two models have distinct predictions (Fig.~\ref{fig:lchc}C). It is only the change in the mean membrane potential that contributes to a shrinking magnitude of correlations in the RG model (Fig.~\ref{fig:orientchange}F, 5B). In the DSP model, one component contributing to contrast-related changes is the increased variance, which causes larger spike count correlations (Fig.~\ref{fig:variancechange}F). The effect of decreased mean seems to be more complex: it results in increased correlations in a wide range of parameters and decreased correlations in a specific subspace of the parameters (Fig.~\ref{fig:corrdep}C). Remarkably, this subspace is characterized by low firing rates and relatively high Fano factors. In summary, contrast modulation related changes in membrane potential variance introduce changes in Fano factors and spike count correlations in both models. The magnitude of the variance change determines the difference in Fano factors between high contrast and low contrast conditions but the direction of deviation is the same for both models. Contrast has opposing effects on spike count correlations in the two models. In the DSP model both increased membrane potential variance and decreased membrane potential mean incur higher spike count correlations at lower contrast levels. In the RG model, however, excess variance does not affect spike count correlations and therefore changes in spike count correlations are solely determined by changes in mean membrane potential which ultimately results in decreasing  correlations with decreasing contrast.

\subsection{Contrast- and orientation-dependent modulation of spiking correlations in a population of simple cells}

In order to test the predictions of the two models against experimental data, we simulated the activity of a population of V1 simple cells in response to changes in stimulus. A critical motivation for using population-level analyses instead of the analysis of pairs of neurons is that pairwise analysis of response statistics requires much more precise control of the recording conditions to curb confounding factors. 

As demonstrated by our analysis of the two-neuron model, stimulus change-related changes in private variability do not distinguish between the DSP and RG models. Therefore it is expected that population distributions of Fano factor are indistinguishable too. Spike count Fano factors show little dependence on stimulus orientation and are increased when stimulus contrast is lowered (data not shown), as predicted by both spiking models. As a consequence, we focus on the analysis of stimulus-dependence of correlated variability in spiking responses of a population of V1 neurons.

\begin{figure}
  \begin{center}
    \includegraphics[width=\textwidth]{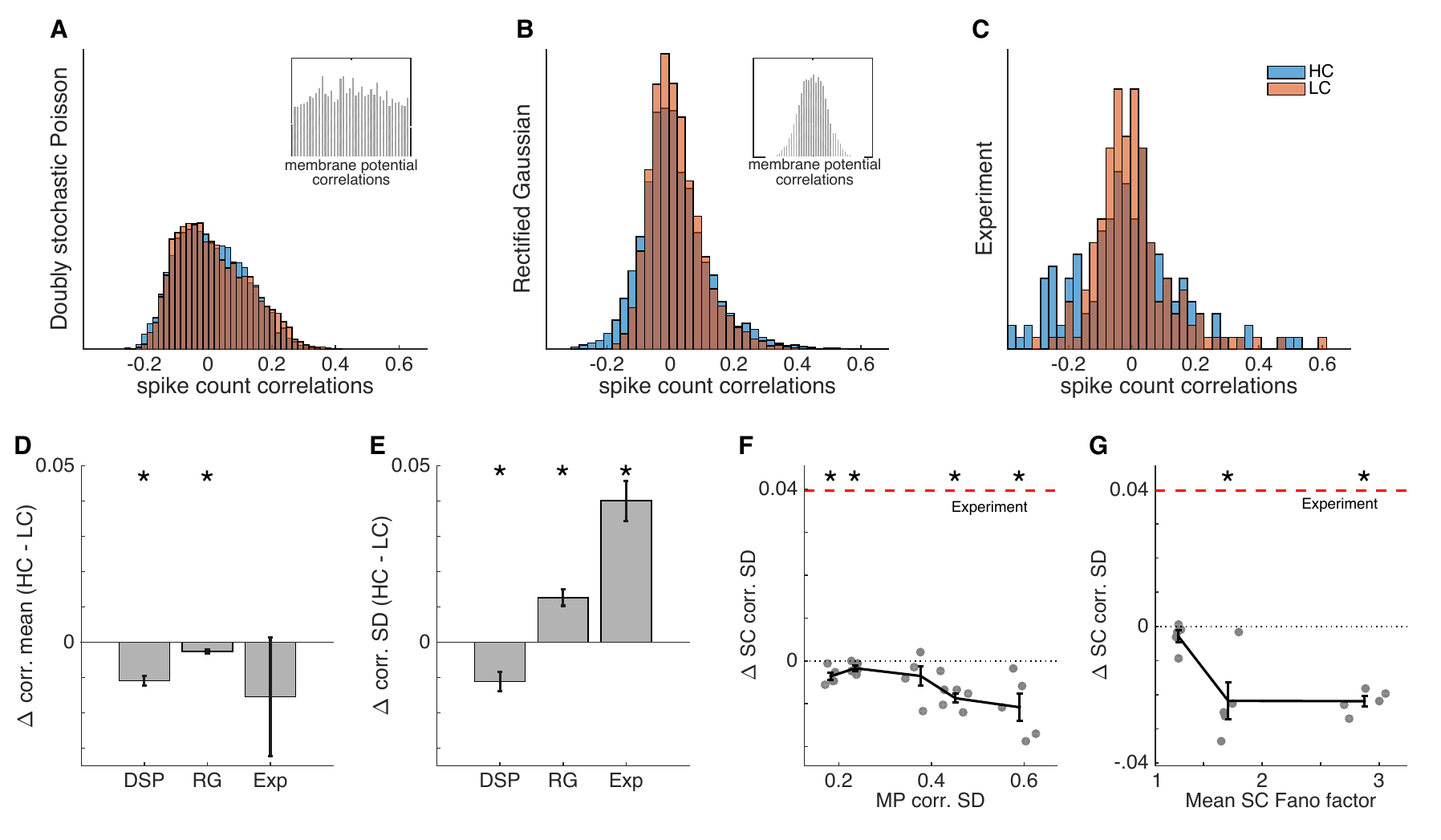}
  \end{center}
  \caption{\textbf{Comparison of model predictions to experimental data - effect of stimulus contrast.} \textbf{\textit{A}}, Distributions of pairwise spike count correlations in a simulated population of 100 DSP neurons in response to a low contrast (LC) and high contrast (HC) stimulus, using membrane potentials (MP) from tuning curves proportional to the contrast. \textbf{\textit{B}}, Distributions of pairwise spike count correlations in a simulated population of 100 RG neurons in response to the same stimuli. \textbf{\textit{C}}, Distributions of measured spike count correlations in response to grating stimuli of low and high contrasts, from \citep{Ecker:2010dn}. \textbf{\textit{D}}, Differences in the means of the two correlation distributions in the two simulated and experimentally recorded populations. Stars represent significant difference of the mean from zero at the p=0.05 level. \textbf{\textit{E}}, Differences in the standard deviations (SD) of the two correlation distributions obtained under the high and low contrast condition. Mean of the differences are shown for the two models across different populations and for experimental recordings across different sets of left-out neurons. The SD mean is significantly different from zero in all populations, with the RG agreeing with data regarding the change of direction and the DSP not. \textbf{\textit{F}}, Dependence of the SD difference in the DSP model on the width of MP correlation distribution. \textbf{\textit{G}}, SD difference in the DSP model at different values of MP mean and variance, shown as a function of the spike count Fano factor, testing whether low mean and high variance in MP can reverse contrast effects in spike count correlations.}
  \label{fig:contrast}
\end{figure}

First, we tested how decreasing stimulus contrast affects the population distribution of spike count correlations in the two models, and compared the results to experimental  data. We set up two populations of 100 model neurons each, implementing different spiking profiles corresponding to the DSP and the RG models. We simulated membrane potential using tuning curve responses to a full-field grating stimulus (see Materials and Methods). Single-cell spiking statistics in the model populations were matched (mean firing rates were 6.5 Hz and 6.3 Hz, mean Fano factors were and 1.3 and 1.2 in the high contrast condition in the DSP and RG populations, respectively). The membrane potential correlations in the two populations were chosen such that mean and width of the two spike count correlation distributions are matched at high contrast stimulus presentation (Fig 8A-B, insets). The distribution of spike count correlation was tuned to have width and mean comparable to physiological values observed in the data recorded (-0.02 and 0.16 for the mean and standard deviation in high contrast conditions, respectively) \citep{Ecker:2010dn}. Width of the distribution was wider for membrane potentials than spike counts both in the case of the DSP and RG models and, as expected from previous analyses, the width of the membrane potential correlation was wider for the DSP model (0.56 and 0.23 for the DSP and RG models, respectively). Lowering the stimulus contrast made the spike count correlation distribution wider in the DSP population (Fig.~\ref{fig:contrast}A). This is caused by the same phenomenon as the increase in pairwise correlation from higher to lower contrast (Fig.~\ref{fig:lchc}C), namely that lower firing rates and increased membrane potential variance tilts the balance between correlated membrane potential and independent spiking variability towards the former. Conversely, in the RG population, the correlation distribution gets narrower with decreasing stimulus contrast (Fig.~\ref{fig:contrast}B). This effect is again in agreement with the results on  pairwise correlation (Fig.~\ref{fig:lchc}C). 

Analyzing population responses from V1 reveals contrast-dependent changes in the distribution of spike count correlations. The distribution of spike count correlations is narrower in response to low contrast stimuli than in response to high contrast (standard deviations of 0.16 for high and 0.12 for low contrast, respectively, Fig.~\ref{fig:contrast}C). The uncertainty of these estimates was assessed by bootstrapping (discarding 20\% of the pairs 5 times), providing an estimate for the standard error of mean for the change in the standard deviation of the spike count correlation distribution. The same analysis was performed for both of the models to obtain a comparable estimate of uncertainty using 5 simulations using 41 randomly selected units, similarly to the number of units available from the experiment. Shrinking of the width of the distribution was significant (Fig.~\ref{fig:contrast}E, one-sample t-test t(4) = 7.06, p = 0.002), similar to the shrinking of the width of the distribution of correlations in the RG model (standard deviation of 0.11 and 0.08 in the high and low contrast conditions respectively, one-sample t-test t(4) = 5.51, p = 0.005, Fig 8E). This result is in contrast with the increased width of spike count correlation distribution at the DSP model (standard deviations of 0.106 and 0.111 in the high and low contrast conditions respectively, one-sample t-test t(4) = -4.06, p = 0.015, Fig.~\ref{fig:contrast}E). Changes in the mean of correlations are relatively small and are consistent across both models and experimental data (in the experimental population, -0.015 and 0.001 in the high and low contrast condition respectively, one-sample t-test t(4) = -0.92, p = 0.41; in the DSP population, 0.009 and 0.015, one-sample t-test t(4) = -7.7, p = 0.002; in the RG population, 0.013 and 0.017, one-sample t-test t(4) = -4.5, p = 0.01; Fig.~\ref{fig:contrast}D). 

In order to test the robustness of the dissociation of the two models by contrast related changes in spike count correlations, we explored how much the changes in correlation distributions depend on the specific settings we used in the model. In the case of the RG model, intuitions obtained from the two-neuron analysis confirm that the increased width is robust against changes in parameters. In the case of the DSP model, however, the specific settings might affect the direction of the change (see Fig.~\ref{fig:corrdep}B,C). We varied the width of the membrane potential correlation distribution for the DSP model to assess whether the contrast dependency of spike count correlations could be reversed at specific settings. Simulations confirmed that this manipulation is not capable of reproducing the experimentally observed pattern (the changes being significantly or non-significantly negative at all tested values, Fig.~\ref{fig:contrast}F). The analysis described in Fig.~\ref{fig:corrdep} motivated us to investigate whether the difference between spike count correlation widths in the high and low contrast conditions could also be positive in the DSP model population when intensively decreasing means and increasing variances concomitantly in the membrane potential. Such changes did not produce a narrowing spike count distribution in response to lower contrast stimuli within (and neither way above) the Fano factor range plausibly observed in measurements (the changes being significantly or non-significantly negative at all tested values, Fig.~\ref{fig:contrast}G). These analyses highlight that the population response statistics to contrast-varied stimuli reliably discriminate between the DSP and RG models of neural spiking and it is the RG model that provides predictions compatible with contrast change related data. 

\begin{figure}
  \begin{center}
    \includegraphics[width=\textwidth]{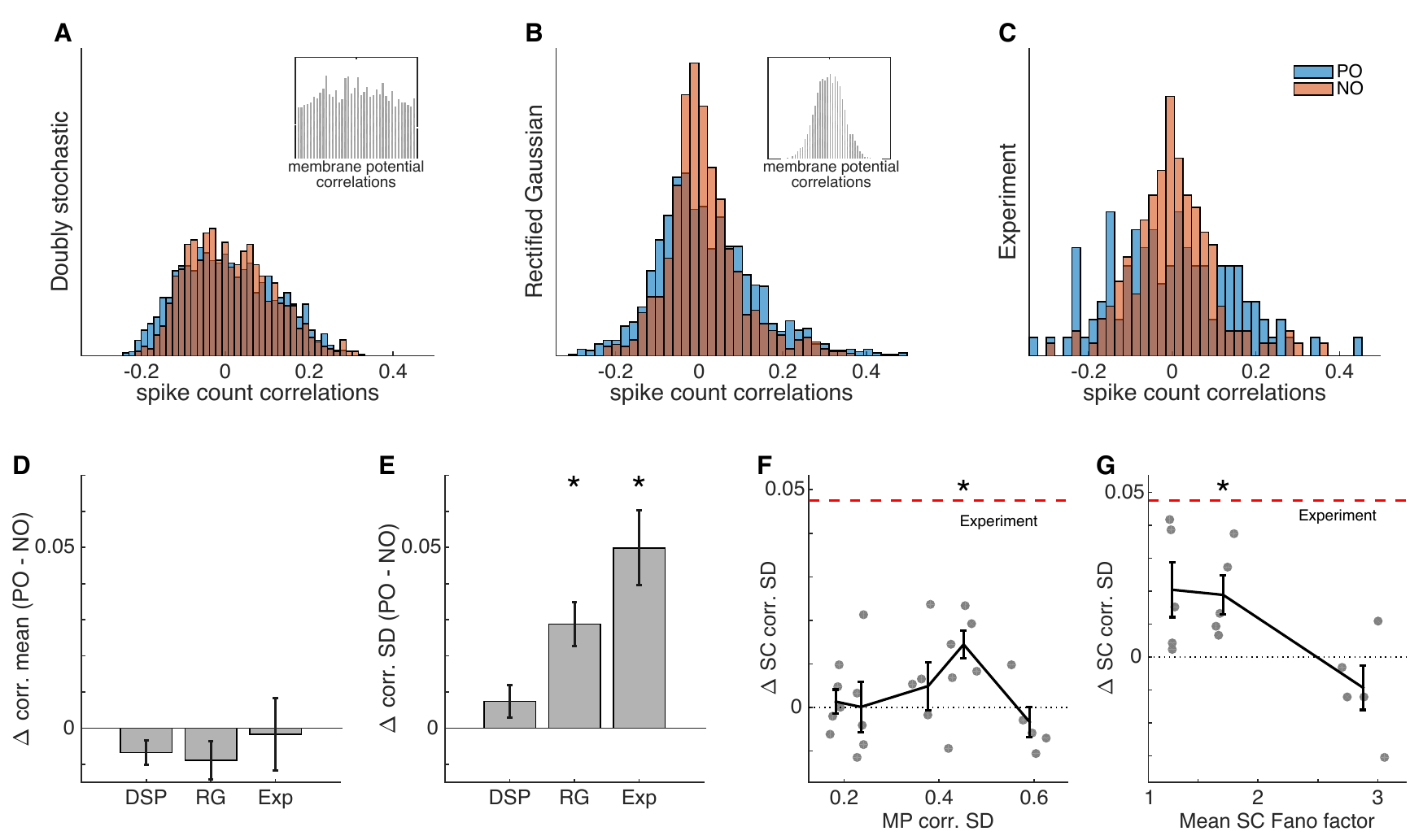}
  \end{center}
  \caption{\textbf{Comparison of model predictions to experimental data - effect of stimulus orientation.} \textbf{\textit{A}}, Distributions of pairwise spike count correlations in a simulated population of 100 DS neurons from pairs observing stimuli of preferred orientation (PO) and non-preferred orientation (NO), using membrane potentials (MP) from tuning curves distributed uniformly over orientations. \textbf{\textit{B}}, Distributions of pairwise spike count correlations in a simulated population of 100 RG neurons in response to the same stimuli. \textbf{\textit{C}}, Distributions of measured spike count correlations in response to grating stimuli of different orientations, from \citep{Ecker:2010dn}. \textbf{\textit{D}}, Differences in the means of the two correlation distributions in the two simulated and the measured populations. Stars represent significant difference of the mean from zero at the 0.05 level. \textbf{\textit{E}}, Differences in the standard deviations (SD) of the two correlation distributions obtained from pairs of neurons observing stimuli with preferred or non-preferred orientations. Mean of the differences are shown for the two models across different populations and for experimental recordings across different sets of left-out neurons. The SD mean is significantly different from zero in the experimental and the RG populations, but not in the DSP population. \textbf{\textit{F}}, Dependence of the SD difference in the DSP model on the width of MP correlation distribution. \textbf{\textit{G}}, SD difference in the DSP model at different values of MP mean and variance, shown as a function of the spike count Fano factor, testing whether low mean and high variance in MP can reproduce experimentally measured orientation effects in spike count correlations.}
  \label{fig:orient}
\end{figure}

Next, we analyzed population responses to changes in stimulus orientation. Population models are constructed similarly to the analysis of contrast dependence. Preferred orientations are uniformly distributed, therefore a simple orientation change is not expected to cause changes in population response statistics. In order to assess orientation related changes in population responses, we classified any particular orientation as preferred or non-preferred orientation based on whether the response of the neuron was above or below its average response intensity. In order to construct distribution of correlations for preferred and non-preferred directions, pairs of cells were selected based on whether both of the cells had the actual stimulus among their preferred orientation or both had it among their non-preferred orientations. Simulation results were contrasted with recordings of populations of V1 units \citep{Ecker:2010dn} in which we also separated pairs of units observing preferred and non-preferred stimuli (see Materials and Methods). In the experimental dataset, we observed a narrower distribution of spike count correlations in response to non-preferred stimuli compared to the response to preferred ones (standard deviations of 0.15 for preferred and 0.10, for non-preferred stimuli, one-paired t-test t(4)=4.83, p=0.009, Fig.~\ref{fig:orient}C,E). In the DSP population, the width of the spike count correlation distribution did not change significantly between the two conditions (standard deviations of 0.11 and 0.1, one-sample t-test t(4)=1.65, p=0.17, Fig.~\ref{fig:orient}A,E). In the RG population, the non-preferred stimulus elicited a narrower distribution of correlations than the preferred one (standard deviations of 0.12 and 0.09, one-sample t-test t(4)=4.77, p=0.009, Fig.~\ref{fig:orient}B,E). Similar to the simulations and experimental data of contrast dependence of responses, changes in the means of correlation distributions were small (in the experimental population, -0.001 and 0.001 in the preferred and non-preferred condition respectively, one-sample t-test t(4) = -0.16, p = 0.88; in the DSP population, 0.006 and 0.011, one-sample t-test t(4) = -2, p = 0.12; in the RG population, 0.016 and 0.01, one-sample t-test t(4) = -1.7, p = 0.17; Fig.~\ref{fig:orient}D). Analysis of the membrane potential parameter space revealed that most correlation, mean and variance values yield non-significant changes between the spike count correlation widths in the preferred and non-preferred conditions, with some small, significantly positive values. (Fig.~\ref{fig:orient}F-G). While experimental results do not provide a direct dissociation between the two models in terms of predictions related to the dependence of correlations on orientation preference, they are reproduced without any particular tuning of model parameters in the case of the RG model, but the DSP model can only account for the patterns in experimental data with specific parameter tuning.

\section{Discussion}

We analyzed widely used models of spiking to investigate their power to predict stimulus-dependent changes in not only in single-cell statistics but in joint statistics of activity too.  The Doubly Stochastic Poisson model \citep{Gur:1997ur} assumes stochastic spiking to account for linear scaling of spike count variance with spike count mean and relies on a separate stochastic process to model the covariance structure of spiking responses. The Rectified Gaussian model \citep{Carandini:2004ee} assumes a single source of stochasticity at the level of membrane potentials but relies on a quasi-deterministic process of spike generation. We demonstrated that while in terms of single-cell statistical measures the models make similar predictions for stimulus change-related modulations in response statistics, predictions on pairwise correlations are distinct. Using a model of a pair of neurons, simulated changes in stimulus orientation and stimulus contrast revealed opposing changes in spike count correlations. The key intuition behind this finding is that the level of spike count correlation with a given level of membrane potential correlation is determined by two phenomena: 1, higher mean membrane potential implies higher magnitude spike count correlations \citep{deLaRocha:2007go} and 2, higher firing rate implies lower level of spike count correlations when Poisson spiking stochasticity is present. The interaction of these two processes result in opposing changes in the correlation structure of the population. In order to assess which of the two models is compatible with neural recordings, response statistics of a population of neurons was simulated in both models and contrasted with response statistics of neurons recorded in V1 of awake monkeys. We have shown that the predictions of the Rectified Gaussian model are in line with electrophysiological data while the Doubly Stochastic Poisson model is not capable of reproducing the patterns of noise correlations in V1 neurons. Our analyses highlight the necessity to assess joint responses of neurons when constructing models of population activity and also highlight that assessment of the width of the distribution of the correlations beyond the mean of correlations can be an important factor. These analyses provide important constraints on the models that can be used effectively to characterize computations in neural populations. 

In our analysis we adopted an approach where we focused on matching the spike count statistics of the models without particular emphasis on the interpretation of the actual levels of  membrane potentials. This approach is motivated by the aim to directly contrast changes in spiking statistics between the two models. It is obvious that separate tuning of the membrane potential parameters could result in membrane potential values that are easier to interpret physiologically. Nonetheless, the assumptions of the DSP model necessitate choices that are hard to reconcile with neuronal data. For instance, in the analysis of the pair of neurons we used a membrane potential correlation of 0.95 in the DSP model in order to obtain a spike count correlation level matching that of the RG model (approximately 0.1) when firing rate was about 35 Hz. In the population model of spiking statistics less extreme values were used, still the tendency of the DSP model to wash out correlations necessitated a wider correlation distribution for the DSP than for the RG model. 

Recent studies using approaches close to our DSP model \citep{Ecker:2016hc, Rabinowitz:2015fa} discuss top-down modulation of response correlations in contrast with the bottom-up modulations discussed here. The important contribution of these models is that different forms of correlations in spiking responses are explained based on simple computational principles. The differences between those approaches and ours are important and can be instructive for future work. First, correlations are introduced at the level of firing rates instead of membrane potentials. Since the variability is introduced prior to spike generation, the conclusions of those studies can be easily translated to our approach solely by assessing the effect of the firing rate nonlinearity. Second, these studies used a single scalar (or a low dimensional) stochastic variable to model gain modulation of visual cortical neurons. The collective gain modulation implies a firing rate correlation of one between those neurons that share the modulatory signal. In our case correlated activity was introduced by a multivariate stochastic process which implements a softer coupling between neurons. Nevertheless, the analysis of the DSP model reveals that realistic firing rates, Fano factors, and spike count correlations  require a surprisingly high level of membrane potential correlation. Third, the use of Poisson stochasticity in these models implies that only variability beyond the Poisson variability are meant to be accounted for. Our analysis shows that the RG model can be a more effective model for variability than the DSP. Since the RG model only assumes a single source of variability at the level of membrane potentials, it provides an opportunity to account for a larger portion of variability and therefore provides an opportunity for models with better predictive power.

We chose to use independent temporal evolution for membrane potentials, which meant independent membrane potential samples in fixed, short time intervals. This was motivated by the relatively fast decay of membrane potential correlations \citep{Azouz:1999wj}. This treatment explicitly incorporates within-trial variability to model the shared variability of neurons. This is in contrast with other models of response variability \citep{Ecker:2016hc}, which consider processes that produce trial-to-trial variability since changes in attentional modulation occur on a slower time scale. We argued in the paper that spiking statistics provide constraints on the spiking models, therefore we believe that patterns in the auto-, and cross-correlation functions \citep{Smith:2008gv} provide further constraints on neural models of spiking. 

Poisson-like firing has been used extensively in the literature \citep{Goris:2015hp, Ma:2006bh, Jazayeri:2006fk, Ma:2014in, Simoncelli:2004ue, Pillow:2007wh, Froudarakis:2014fs}. Besides its capability to provide a parsimonious explanation of the relationship between spiking intensity and spiking variability, Poisson neurons have much theoretical appeal too. First, Poisson-like spiking distribution ensures that fitting network parameters is a convex optimization problem, or in other words, there is a single (global) maximum in optimization \citep{Paninski:2004to}. Second, in theories of encoding information via populations of neurons, a Poisson-like likelihood function provides a representation in which the log likelihood contains linear terms which enables simple, neurally plausible computations \citep{Ma:2006bh, Jazayeri:2006fk}. In contrast with models based on the Poisson assumption, alternative approaches have applied models that have a closer relationship to the Rectified Gaussian model \citep{Paninski:2004jc, Brette:2005ke, Lin:2015dw} and argued for the capability of such models to predict both for single-cell response properties \citep{Brette:2005ke} and population statistics \citep{Lin:2015dw}. Response variability in these two model classes is approached in two markedly different ways: in one, stochasticity is part of the spike generation process, in the other it originates prior to spike generation and can be related to membrane potential-level processes. In this context, our study contributes to the field by a direct and controlled comparison of the predictions of these approaches on spiking statistics. 

Recent advances in modelling data recorded from a large number of neurons have helped to assess neural responses in a trial-by-trial manner and to relate them to variances in behavior \citep{Churchland:2010he, Yu:2009ex}, disentangle mixed sensitivities \citep{Kobak:2016eo}, eliminate noise by tracing neural variability back to changes in latent factors \citep{Machens:2010in} and to implement closed-loop brain-computer interfacing \citep{Sadtler:2015kx}. Modelling and predicting correlations is a critical factor in population-level analyses of neuronal data \citep{Cunningham:2014ev}. Those are precisely the correlated changes in neuronal activity that help to eliminate the effect of uncontrolled variables, to reduce apparent noise in the measurements, and to predict the activity of missing neurons. Therefore, it is crucial to have an adequate model of response variability that can predict the effects of changes in stimulus on the correlation structure. The stochasticity assumed to underlie observed spiking variability can take on the form of Poisson variability \citep{Archer:2014wn, Macke:2011ut} or Gaussian noise \citep{Yu:2009ex, Sadtler:2015kx}. Our study aims to provide constraints on the forms of stochasticity in these models by emphasising that bottom-up driven changes in the response statistics can differentiate between alternative models. As demonstrated in the paper, even when single-cell response statistics have limited power to distinguish between alternative models, joint statistics can reveal properties that are incompatible with the predictions of one or the other. We expect that proper understanding and characterization of stochasticity in the nervous system helps to better interpret joint statistics and especially correlations present in the activity of neural populations.

\section*{Acknowledgements}

We thank M. Lengyel for useful discussions, D. G. Nagy for comments on the manuscript and especially A. Ecker, P. Berens, M. Bethge and A. Tolias for making their data publicly available. This work was supported by a Lend\"ulet Award of the Hungarian Academy of Sciences (G.O., M.B.) and an award from the National Brain Research Program of Hungary (NAP-B). The authors declare no competing financial interests.

\bibliographystyle{apalike}
\bibliography{bib2,bib}

\begin{thebibliography}{}

\bibitem[Archer et~al., 2014]{Archer:2014wn}
Archer, E.~W., Koster, U., Pillow, J.~W., and Macke, J.~H. (2014).
\newblock {Low-dimensional models of neural population activity in sensory
  cortical circuits}.
\newblock {\em Advances in Neural Information Processing Systems}, pages
  343--351.

\bibitem[Azouz and Gray, 1999]{Azouz:1999wj}
Azouz, R. and Gray, C.~M. (1999).
\newblock {Cellular mechanisms contributing to response variability of cortical
  neurons in vivo.}
\newblock {\em The Journal of Neuroscience}, 19(6):2209--2223.

\bibitem[Brette and Gerstner, 2005]{Brette:2005ke}
Brette, R. and Gerstner, W. (2005).
\newblock {Adaptive exponential integrate-and-fire model as an effective
  description of neuronal activity.}
\newblock {\em Journal of Neurophysiology}, 94(5):3637--3642.

\bibitem[Britten et~al., 1993]{Britten:1993wv}
Britten, K.~H., Shadlen, M.~N., Newsome, W.~T., and Movshon, J.~a. (1993).
\newblock {Responses of neurons in macaque MT to stochastic motion signals.}
\newblock {\em Visual Neuroscience}, 10(6):1157--1169.

\bibitem[Carandini, 2004]{Carandini:2004ee}
Carandini, M. (2004).
\newblock {Amplification of trial-to-trial response variability by neurons in
  visual cortex.}
\newblock {\em PLoS Biology}, 2(9):E264.

\bibitem[Carandini and Ferster, 2000]{Carandini:2000ud}
Carandini, M. and Ferster, D. (2000).
\newblock {Membrane potential and firing rate in cat primary visual cortex.}
\newblock {\em The Journal of Neuroscience}, 20(1):470--484.

\bibitem[Churchland et~al., 2011]{Churchland:2011hd}
Churchland, A.~K., Kiani, R., Chaudhuri, R., Wang, X.~J., Pouget, A., and
  Shadlen, M.~N. (2011).
\newblock {Variance as a Signature of Neural Computations during Decision
  Making}.
\newblock {\em Neuron}, 69(4):818--831.

\bibitem[Churchland et~al., 2010]{Churchland:2010he}
Churchland, M.~M., Yu, B.~M., Cunningham, J.~P., Sugrue, L.~P., Cohen, M.~R.,
  Corrado, G.~S., Newsome, W.~T., Clark, A.~M., Hosseini, P., Scott, B.~B.,
  Bradley, D.~C., Smith, M.~a., Kohn, A., Movshon, J.~A., Armstrong, K.~M.,
  Moore, T., Chang, S.~W., Snyder, L.~H., Lisberger, S.~G., Priebe, N.~J.,
  Finn, I.~M., Ferster, D., Ryu, S.~I., Santhanam, G., Sahani, M., and Shenoy,
  K.~V. (2010).
\newblock {Stimulus onset quenches neural variability: a widespread cortical
  phenomenon.}
\newblock {\em Nature Neuroscience}, 13(3):369--378.

\bibitem[Cohen and Kohn, 2011]{Cohen:2011eh}
Cohen, M.~R. and Kohn, A. (2011).
\newblock {Measuring and interpreting neuronal correlations}.
\newblock {\em Nature Neuroscience}, 14(7):811--819.

\bibitem[Cunningham and Yu, 2014]{Cunningham:2014ev}
Cunningham, J.~P. and Yu, B.~M. (2014).
\newblock {Dimensionality reduction for large-scale neural recordings}.
\newblock {\em Nature Neuroscience}, 17(11):1500--1509.

\bibitem[de~la Rocha et~al., 2007]{deLaRocha:2007go}
de~la Rocha, J., Doiron, B., Shea-Brown, E., Josi~c, K. s.~i., and Reyes, A.
  (2007).
\newblock {Correlation between neural spike trains increases with firing rate}.
\newblock {\em Nature}, 448(7155):802--806.

\bibitem[Ecker et~al., 2014]{Ecker:2014cl}
Ecker, A.~S., Berens, P., Cotton, R.~J., Subramaniyan, M., Denfield, G.~H.,
  Cadwell, C.~R., Smirnakis, S.~M., Bethge, M., and Tolias, A.~S. (2014).
\newblock {State dependence of noise correlations in macaque primary visual
  cortex}.
\newblock {\em Neuron}, 82(1):235--248.

\bibitem[Ecker et~al., 2010]{Ecker:2010dn}
Ecker, A.~S., Berens, P., Keliris, G.~A., Bethge, M., Logothetis, N.~K., and
  Tolias, A.~S. (2010).
\newblock {Decorrelated Neuronal Firing in Cortical Microcircuits}.
\newblock {\em Science}, 327(5965):584--587.

\bibitem[Ecker et~al., 2016]{Ecker:2016hc}
Ecker, A.~S., Denfield, G.~H., Bethge, M., and Tolias, A.~S. (2016).
\newblock {On the Structure of Neuronal Population Activity under Fluctuations
  in Attentional State}.
\newblock {\em Journal of Neuroscience}, 36(5):1775--1789.

\bibitem[Faisal et~al., 2008]{Faisal:2008cp}
Faisal, A.~A., Selen, L. P.~J., and Wolpert, D.~M. (2008).
\newblock {Noise in the nervous system.}
\newblock {\em Nature Reviews Neuroscience}, 9(4):292--303.

\bibitem[Finn et~al., 2007]{Finn:2007hc}
Finn, I.~M., Priebe, N.~J., and Ferster, D. (2007).
\newblock {The Emergence of Contrast-Invariant Orientation Tuning in Simple
  Cells of Cat Visual Cortex}.
\newblock {\em Neuron}, 54(1):137--152.

\bibitem[Froudarakis et~al., 2014]{Froudarakis:2014fs}
Froudarakis, E., Berens, P., Ecker, A.~S., Cotton, R.~J., Sinz, F.~H.,
  Yatsenko, D., Saggau, P., Bethge, M., and Tolias, A.~S. (2014).
\newblock {Population code in mouse V1 facilitates readout of natural scenes
  through increased sparseness}.
\newblock {\em Nature Neuroscience}, 17(6):851--857.

\bibitem[Goris et~al., 2014]{Goris:2014jg}
Goris, R. L.~T., Movshon, J.~A., and Simoncelli, E.~P. (2014).
\newblock {Partitioning neuronal variability}.
\newblock {\em Nature Neuroscience}, 17(6):858--865.

\bibitem[Goris et~al., 2015]{Goris:2015hp}
Goris, R. L.~T., Simoncelli, E.~P., and Movshon, J.~A. (2015).
\newblock {Origin and Function of Tuning Diversity in Macaque Visual Cortex}.
\newblock {\em Neuron}, pages 1--14.

\bibitem[Gur et~al., 1997]{Gur:1997ur}
Gur, M., Beylin, A., and Snodderly, D.~M. (1997).
\newblock {Response variability of neurons in primary visual cortex (V1) of
  alert monkeys.}
\newblock {\em Journal of Neuroscience}, 17(8):2914--2920.

\bibitem[Haefner et~al., 2016]{Haefner:2016vr}
Haefner, R.~M., Berkes, P., and Fiser, J. (2016).
\newblock {Perceptual Decision-Making as Probabilistic Inference by Neural
  Sampling.}
\newblock {\em Neuron}, 90(3):649--660.

\bibitem[Haefner et~al., 2013]{Haefner:2013gh}
Haefner, R.~M., Gerwinn, S., Macke, J.~H., and Bethge, M. (2013).
\newblock {Inferring decoding strategies from choice probabilities in the
  presence of correlated variability.}
\newblock {\em Nature Neuroscience}, 16(2):235--242.

\bibitem[Jazayeri and Movshon, 2006]{Jazayeri:2006fk}
Jazayeri, M. and Movshon, J.~A. (2006).
\newblock {Optimal representation of sensory information by neural
  populations.}
\newblock {\em Nature Neuroscience}, 9(5):690--696.

\bibitem[Karklin and Lewicki, 2009]{Karklin:2009hl}
Karklin, Y. and Lewicki, M.~S. (2009).
\newblock {Emergence of complex cell properties by learning to generalize in
  natural scenes.}
\newblock {\em Nature}, 457(7225):83--86.

\bibitem[Kobak et~al., 2016]{Kobak:2016eo}
Kobak, D., Brendel, W., Constantinidis, C., Feierstein, C.~E., Kepecs, A.,
  Mainen, Z.~F., Romo, R., Qi, X.-L., Uchida, N., and Machens, C.~K. (2016).
\newblock {Demixed principal component analysis of neural population data}.
\newblock {\em eLife}, 5:e10989.

\bibitem[Kohn and Smith, 2005]{Kohn:2005um}
Kohn, A. and Smith, M.~a. (2005).
\newblock {Stimulus dependence of neuronal correlation in primary visual cortex
  of the macaque.}
\newblock {\em Journal of Neuroscience}, 25(14):3661--3673.

\bibitem[Lewandowski et~al., 2009]{Lewandowski:2009gd}
Lewandowski, D., Kurowicka, D., and Joe, H. (2009).
\newblock {Generating random correlation matrices based on vines and extended
  onion method}.
\newblock {\em Journal of Multivariate Analysis}, 100(9):1989--2001.

\bibitem[Lin et~al., 2015]{Lin:2015dw}
Lin, I.-C., Okun, M., Carandini, M., and Harris, K.~D. (2015).
\newblock {The Nature of Shared Cortical Variability}.
\newblock {\em Neuron}, 87(3):644--656.

\bibitem[Ma et~al., 2006]{Ma:2006bh}
Ma, W.~J., Beck, J.~M., Latham, P.~E., and Pouget, A. (2006).
\newblock {Bayesian inference with probabilistic population codes.}
\newblock {\em Nature Neuroscience}, 9(11):1432--1438.

\bibitem[Ma and Jazayeri, 2014]{Ma:2014in}
Ma, W.~J. and Jazayeri, M. (2014).
\newblock {Neural Coding of Uncertainty and Probability}.
\newblock {\em Annual Review of Neuroscience}, 37(1):205--220.

\bibitem[Machens et~al., 2010]{Machens:2010in}
Machens, C.~K., Romo, R., and Brody, C.~D. (2010).
\newblock {Functional, But Not Anatomical, Separation of "What" and "When" in
  Prefrontal Cortex}.
\newblock {\em Journal of Neuroscience}, 30(1):350--360.

\bibitem[Macke et~al., 2011]{Macke:2011ut}
Macke, J.~H., Buesing, L., Cunningham, J.~P., Yu, B.~M., Shenoy, K.~V., and
  Sahani, M. (2011).
\newblock {Empirical models of spiking in neural populations}.
\newblock {\em Advances in Neural Information Processing Systems}, pages
  1350--1358.

\bibitem[Mainen and Sejnowski, 1995]{Mainen:1995uz}
Mainen, Z.~F. and Sejnowski, T.~J. (1995).
\newblock {Reliability of spike timing in neocortical neurons.}
\newblock {\em Science}, 268(5216):1503--1506.

\bibitem[Moreno-Bote et~al., 2008]{MorenoBote:2008gg}
Moreno-Bote, R. e.~n., Renart, A., and Parga, N. e.~s. (2008).
\newblock {Theory of input spike auto- and cross-correlations and their effect
  on the response of spiking neurons.}
\newblock {\em Neural Computation}, 20:1651--1705.

\bibitem[Paninski, 2004]{Paninski:2004to}
Paninski, L. (2004).
\newblock {Maximum likelihood estimation of cascade point-process neural
  encoding models.}
\newblock {\em Network: Computation in Neural Systems}, 15(4):243--262.

\bibitem[Paninski et~al., 2004]{Paninski:2004jc}
Paninski, L., Pillow, J.~W., and Simoncelli, E.~P. (2004).
\newblock {Maximum likelihood estimation of a stochastic integrate-and-fire
  neural encoding model.}
\newblock {\em Neural Computation}, 16(12):2533--2561.

\bibitem[Pillow, 2007]{Pillow:2007wh}
Pillow, J.~W. (2007).
\newblock {Likelihood-based approaches to modeling the neural code}.
\newblock {\em Bayesian brain: Probabilistic approaches to neural coding},
  pages 53--70.

\bibitem[Rabinowitz et~al., 2015]{Rabinowitz:2015fa}
Rabinowitz, N.~C., Goris, R.~L., Cohen, M., and Simoncelli, E.~P. (2015).
\newblock {Attention stabilizes the shared gain of V4 populations.}
\newblock {\em eLife}, 4:e08998.

\bibitem[Renart and Machens, 2014]{Renart:2014dr}
Renart, A. and Machens, C.~K. (2014).
\newblock {Variability in neural activity and behavior}.
\newblock {\em Neural Computation}, 25:211--220.

\bibitem[Ruff and Cohen, 2014]{Ruff:2014fa}
Ruff, D.~A. and Cohen, M.~R. (2014).
\newblock {Attention can either increase or decrease spike count correlations
  in visual cortex.}
\newblock {\em Nature Neuroscience}, 17(11):1591--1597.

\bibitem[Sadtler et~al., 2015]{Sadtler:2015kx}
Sadtler, P.~T., Quick, K.~M., Golub, M.~D., Chase, S.~M., Ryu, S.~I.,
  Tyler-Kabara, E.~C., Yu, B.~M., and Batista, A.~P. (2015).
\newblock {Neural constraints on learning}.
\newblock {\em Nature}, 512(7515):423--426.

\bibitem[Simoncelli et~al., 2004]{Simoncelli:2004ue}
Simoncelli, E.~P., Paninski, L., Pillow, J., and Schwartz, O. (2004).
\newblock {Characterization of neural responses with stochastic stimuli}.
\newblock In Gazzaniga, M., editor, {\em The cognitive neurosciences }, pages
  327--338. MIT Press , Cambridge, MA.

\bibitem[Smith and Kohn, 2008]{Smith:2008gv}
Smith, M.~a. and Kohn, A. (2008).
\newblock {Spatial and temporal scales of neuronal correlation in primary
  visual cortex.}
\newblock {\em The Journal of Neuroscience}, 28(48):12591--12603.

\bibitem[Softky and Koch, 1993]{Softky:1993uj}
Softky, W.~R. and Koch, C. (1993).
\newblock {The highly irregular firing of cortical cells is inconsistent with
  temporal integration of random EPSPs.}
\newblock {\em The Journal of neuroscience : the official journal of the
  Society for Neuroscience}, 13(1):334--350.

\bibitem[Tolhurst et~al., 1983]{Tolhurst:1983wa}
Tolhurst, D.~J., Movshon, J.~A., and Dean, A.~F. (1983).
\newblock {The statistical reliability of signals in single neurons in cat and
  monkey visual cortex.}
\newblock {\em Vision Research}, 23(8):775--785.

\bibitem[Tolhurst et~al., 1981]{Tolhurst:1981ju}
Tolhurst, D.~J., Movshon, J.~a., and Thompson, I.~D. (1981).
\newblock {The dependence of response amplitude and variance of cat visual
  cortical neurones on stimulus contrast}.
\newblock {\em Experimental Brain Research}, 41:414--419.

\bibitem[Tomko and Crapper, 1974]{Tomko:1974ul}
Tomko, G.~J. and Crapper, D.~R. (1974).
\newblock {Neuronal variability: non-stationary responses to identical visual
  stimuli.}
\newblock {\em Brain research}, 79(3):405--418.

\bibitem[Yu et~al., 2009]{Yu:2009ex}
Yu, B.~M., Cunningham, J.~P., Santhanam, G., Ryu, S.~I., Shenoy, K.~V., and
  Sahani, M. (2009).
\newblock {Gaussian-process factor analysis for low-dimensional single-trial
  analysis of neural population activity.}
\newblock {\em Journal of Neurophysiology}, 102(1):614--635.

\bibitem[Yu and Ferster, 2010]{Yu:2010p314}
Yu, J. and Ferster, D. (2010).
\newblock {Membrane potential synchrony in primary visual cortex during sensory
  stimulation}.
\newblock {\em Neuron}, 68(6):1187--1201.

\end{thebibliography}

\end{document}